\documentclass[acmsmall]{acmart}
\usepackage{subfigure}
\usepackage{times}
\usepackage{epsfig}
\usepackage{graphicx}
\usepackage{amsmath}

\usepackage{multirow}
\usepackage{subfigure}
\usepackage{color}

\AtBeginDocument{%
  \providecommand\BibTeX{{%
    \normalfont B\kern-0.5em{\scshape i\kern-0.25em b}\kern-0.8em\TeX}}}

\begin{document}

\title{Point Cloud Quality Assessment: Dataset Construction and Learning-based No-Reference Metric}


\author{Yipeng~Liu}
\email{liuyipeng@sjtu.edu.cn}
\affiliation{%
  \institution{Cooperative Medianet Innovation Center, Shanghai Jiaotong University}
  \streetaddress{Dongchuan Road 800}
  \state{Shanghai}
  \country{China}
}
\author{Qi~Yang}
\email{yang\_littleqi@sjtu.edu.cn}
\affiliation{%
  \institution{Cooperative Medianet Innovation Center, Shanghai Jiaotong University}
  \streetaddress{Dongchuan Road 800}
  \state{Shanghai}
  \country{China}
}
\author{Yiling~Xu}
\email{yl.xu@sjtu.edu.cn}
\affiliation{%
  \institution{Cooperative Medianet Innovation Center, Shanghai Jiaotong University}
  \streetaddress{Dongchuan Road 800}
  \state{Shanghai}
  \country{China}
}
\author{Le~Yang}
\email{le.yang@canterbury.ac.nz}
\affiliation{%
  \institution{Department of electrical and computer engineering, University of Canterbury}
  \streetaddress{Christchurch 8041}
  \state{Canterbury}
  \country{New Zealand}
}

\thanks{This paper is supported in part by National Key Research and Development Program of China (2018YFE0206700), National Natural Science Foundation of China (61971282, U20A20185). Y. Liu and Q. Yang contributed equally to this work. Corresponding author: Y. Xu}

\begin{abstract}
\par Full-reference (FR) point cloud quality assessment (PCQA) has achieved impressive progress in recent years. However, in many cases, obtaining the reference point clouds is difficult, so no-reference (NR) metrics have become a research hotspot. Few researches about NR-PCQA are carried out due to the lack of a large-scale PCQA dataset. In this paper, we first build a large-scale PCQA dataset named LS-PCQA, which includes 104 reference point clouds and more than 22,000 distorted samples. In the dataset, each reference point cloud is augmented with 31 types of impairments (e.g., Gaussian noise, contrast distortion, local missing, and compression loss) at 7 distortion levels. Besides, each distorted point cloud is assigned with a pseudo quality score as its substitute of Mean Opinion Score (MOS). Inspired by the hierarchical perception system and considering the intrinsic attributes of point clouds, we propose a NR metric ResSCNN based on sparse convolutional neural network (CNN) to accurately estimate the subjective quality of point clouds. We conduct several experiments to evaluate the performance of the proposed NR metric. The results demonstrate that ResSCNN exhibits the state-of-the-art (SOTA) performance among all the existing NR-PCQA metrics and even outperforms some FR metrics. The dataset presented in this work will be made publicly accessible at {\url{http://smt.sjtu.edu.cn}}. The source code for the proposed ResSCNN can be found at {\url{https://github.com/lyp22/ResSCNN}}.
\end{abstract}


\maketitle

\section{Introduction}
\par Recently, thanks to the increasing capability of 3D acquisition devices, point cloud has emerged as the most popular format for immersive media. A point cloud consists of a collection of points, each of which has geometric coordinates but may also contain a number of other attributes such as color, reflectance and surface normals. Point cloud has been used in many applications such as augmented reality (AR), autonomous driving, industrial robotics, documentation and facial landmarking~\cite{Demisse2018Facial,Sun2019Landmarking}. In practice, a variety of distortions could be involved and affect human perception. Developing point cloud quality assessment (PCQA) can help to understand the distortions and carry out the quality optimization for distorted point clouds. Generally, PCQA can be performed using subjective experiments or objective metrics. Because the subjective experiment is expensive and time-consuming, studying robust and effective objective PCQA metrics is important. However, different from 2D media, such as images and videos presented in the regular grid, the points of 3D point clouds are scattered in the spatial space. How to extract effective features from scattered points for quality assessment needs further investigation.

\subsection{Motivation}

\par As in image quality assessment (IQA), there are three different types of PCQA metrics, namely, full-reference (FR), reduced-reference (RR) and no-reference (NR) metrics~\cite{Min2021survey,Zhai2021survey}. In FR-PCQA and RR-PCQA, the original reference point cloud or its features are required, while NR-PCQA only utilizes distorted samples. FR-PCQA was first studied and has shown impressive progress in recent years. Some FR-PCQA metrics, including the point-to-point (p2point)~\cite{cignoni1998metro} and point-to-plane (p2plane)~\cite{Mekuria2016Evaluation}, have already been used in the evaluation of the MPEG point cloud compression (PCC) technology. These metrics mainly leverage the point-wise spatial deformation between the reference and distorted samples, which neglects the fact that the human visual system (HVS) is more sensitive to the structural distortions. Moreover, the color-dominating distortions also play an important role in human perception, but these metrics do not take the color-related features into consideration. Thus, metrics considering both the geometry and color distortions have been proposed, such as PCQM~\cite{meynet2020pcmd}, GraphSIM~\cite{yang2020graphsim} and MPED~\cite{yang2021MPED}.

\par However, in many practical scenarios, the reference point cloud is not available. Therefore, developing NR-PCQA metrics can further facilitate the application of point clouds. To the best of our knowledge, few NR-PCQA metrics have been developed yet due to the following reasons:

\par i) Lack of large-scale datasets. The robustness and generalization capability of learning-based NR-PCQA metrics rely on the richness of available data. The lack of a large-scale PCQA dataset impedes the development of NR-PCQA metrics. Unlike 2D images that are relatively easy to collect in a huge volume, obtaining point clouds is far more costly. The existing PCQA datasets, e.g., SJTU-PCQA~\cite{Yang2020TMM3DTO2D}, only have a few hundred point cloud instances, which are not sufficient to derive learning-based NR-PCQA metrics with high generalization ability.

\par ii) Difficulty in data annotation. The labels for PCQA usually come from subjective experiments that are expensive and time-consuming and require strict control conditions. Considering that there are more than 22,000 samples in our newly developed dataset, and each point cloud requires at least 16 subjects after outliers removal to collect Mean Opinion Score (MOS), it is very difficult to annotate the whole dataset via subjective experiments alone.

\par iii) Mismatch between the dense convolution and uneven spatial distribution of scattered points. For learning-based metrics, traditional implementations of the dense convolution are optimized for data on densely populated grids. They are unable to process the uneven disordered data efficiently. The mismatch between the dense convolution and uneven distribution of scattered points results in high memory usage and slow inference speed when using the dense convolution to process point clouds.

\subsection{Our Approach}

\par In this paper, we first build a large-scale PCQA dataset in which the reference  samples have rich geometry and color information, and the distortions lie in both geometry and color domains; secondly, we annotate the built dataset using the pseudo MOS which has been successfully applied in images~\cite{Wu2020PesudoMOS}; thirdly, we propose a sparse convolutional neural network (CNN) based NR-PCQA metric to extract the hierarchical features directly from 3D point clouds, considering both the geometry and texture information.

\par The distortion of point clouds is more complex than that of images. The photometric attribute of point clouds may be subject to similar distortions in 2D images, because the dense point clouds are also produced by optical devices, such as the depth camera and light field camera. Thus, the distortions induced during the image production may also appear in the point clouds, such as Gaussian noise. Besides, the data structure of point clouds leads to some unique distortions, such as the geometrical missing. As a result, the point cloud distortions may lie in their geometry components, or attribute components, or both. To better understand and study the point cloud distortions, we build a PCQA dataset with the largest size so far, which contains 104 reference point clouds with 31 types of distortions (such as Gaussian noise, contrast distortion, local missing and compression loss) at 7 different levels, leading to a total number of more than 22,000 distorted point clouds. Some reference point cloud samples are chosen from~\cite{Mpeg1,Mpeg2,JPEG,Mpeg3,Nouri2017Greyc} which have already been used in the MPEG PCC standards. Some reference point cloud samples are crafted from the mesh format data in~\cite{free3d,sketchfab}.

\par To annotate the newly built dataset, we conduct a large-scale subjective experiment to obtain the subjective MOS for some samples. The experiment is conducted with 1,240 distorted point cloud samples selected from the database covering all distortion types. Then, inspired by~\cite{Wu2020PesudoMOS}, we compute Spearman rank order correlation coefficient (SROCC) of FR-PCQA metrics for each distortion type based on the subjective MOS and their predicted scores. The scores from the best FR-PCQA metric for each distortion type are selected and normalized to obtain pseudo MOS and annotate the whole dataset. Since the performance of some existing FR-PCQA algorithms has been shown to be consistent with the HVS under certain types of distortions, the pseudo MOS can be considered accurate.

\par Finally, considering the uneven distribution of 3D point clouds, the sparse convolution is introduced into NR-PCQA in this work using Minkowski Engine~\cite{Choy2019Minkowski}. We employ a sparse tensor representation and attempt to develop a sparse CNN based NR-PCQA metric called ResSCNN. It extracts the hierarchical features from 3D point clouds using the sparse convolution instead of the traditional dense convolution to avoid the massive increase in the elements of feature map. Besides, the dimensionality reduction techniques, such as down-sampling, are not included in the proposed metric because the dimensionality reduction itself introduces extra distortions.

\par We evaluate the proposed ResSCNN in this paper on the newly established LS-PCQA dataset,  SJTU-PCQA dataset~\cite{Yang2020TMM3DTO2D} and WPC2.0 dataset~\cite{Su2019WPC,Liu2022WPC}. ResSCNN presents robust and competitive performance on all datasets compared with other NR metrics and even some FR metrics.

\subsection{Contributions}
\par The main contributions of this paper are as follows.
\begin{itemize}
\item We establish a large-scale PCQA dataset called LS-PCQA with 22,568 distorted point clouds derived from 104 original reference point clouds, each with 31 types of distortions at 7 distortion levels. The new dataset covers a wide range of impairments during point cloud production, compression, transmission and presentation. To the best of our knowledge, it is the largest PCQA dataset at present.
\item Based on LS-PCQA dataset, we conduct a fairly large subjective experiment to collect MOS. In the experiment, we recruit 224 candidates to score 1,240 distorted point clouds, and ensure that at least 16 valid subjective scores are collected for each distorted point cloud according to ITU-R BT.500~\cite{BT500}. Based on the subjective MOS, pseudo quality scores are calculated to annotate the whole LS-PCQA dataset.
\item We develop a NR-PCQA metric using sparse CNN with only 1.2M parameters. The experiment results show that our proposed metric offers robust and competitive performance compared with other NR metrics and even some FR metrics over three datasets.

  \end{itemize}
\par The rest of this paper is organized as follows. The related work is surveyed in Section~\ref{sec:related work}. The proposed large-scale PCQA dataset, LS-PCQA, is introduced in Section~\ref{sec:dataset}, and Section~\ref{sec:framework} presents the new NR-PCQA metric, ResSCNN, with its performance evaluation given in Section~\ref{sec:experiment}. Finally, the conclusion is drawn in Section~\ref{sec:conclusion}.

\section{Related Work}\label{sec:related work}

\par This section surveys the development of PCQA metrics and 3D feature description.

\subsection{Quality Assessment Metrics}

\par Quality assessment is widely used in images \cite{Jiang2020IQA1,Jiang2018IQA2} and various 3D media formats \cite{Min2020LF,Zhang2021NR1,Zhang2021NR2,Jiang2020SIQA}. For point clouds, some existing quality assessment metrics evaluate the distortion based on geometrical attributes only. Specifically, p2point~\cite{cignoni1998metro} quantifies the distances between corresponding points to measure the degree of distortion. P2plane~\cite{Mekuria2016Evaluation} improves over p2point by projecting the obtained p2point distances along the surface normal direction. The point-to-mesh (p2mesh)~\cite{Tian2017Evaluation} reconstructs the surface and then measures the distance from a point to the surface, but the efficiency of p2mesh is heavily dependent on the accuracy of the surface reconstruction. Both p2point and p2plane have already been applied in the standardized MPEG PCC technology~\cite{MPEGSoft}.

\par On the other hand, Alexiou \emph{et al.} ~\cite{alexiou2018pointt} propose to measure the geometrical distortion based on the angular difference of point normals. Javaheri \emph{et al.}~\cite{javaheri2020haus} propose a generalized Hausdorff distance by employing the $k$th lowest distance instead of the biggest distance to address that Hausdorff distances are over-sensitive to noise.

\par The aforementioned point-wise metrics ignore the fact that HVS is more sensitive to structural features. Besides, color information also plays an important role in PCQA. Considering the huge success of SSIM~\cite{wang2004image} in IQA, researchers start to consider spatial structural features as the quality index. Some of them take geometry and color into consideration simultaneously. Meynet \emph{et al.}~\cite{meynet2019pcmsdm} propose to use the local curvature statistics to reflect the point cloud surface distortion, and further pool curvature and color lightness together via optimally-weighted linear combination~\cite{meynet2020pcmd}. Viola \emph{et al.}~\cite{viol2020acolor} propose a quality assessment metric based on the color histogram.  Alexiou \emph{et al.}~\cite{alexiou2020TowardsStructural} incorporate four types of point cloud attributes, namely, geometry, normal vectors, curvature values and colors, into the form of SSIM~\cite{wang2004image}. Yang \emph{et al.}~\cite{yang2020graphsim} propose to extract point cloud color gradient using graph signal processing to estimate the point cloud quality. Zhang \emph{et al.}~\cite{Zhang2021MSGraphSIM} improve~\cite{yang2020graphsim} using a HVS-based multi-scale method. Javaheri \emph{et al.}~\cite{javaheri2021PTD} propose a point-to-distribution quality assessment metric considering both the geometry and texture.

\par Another idea for PCQA is to project the 3D point cloud into a number of 2D planes, and then the 2D IQA metrics can be used, such as the ones in~\cite{torlig2018novel,alexiou2018pointt,Yang2020TMM3DTO2D,javaheri2021JPC}. However, the selection of projection directions may significantly influence metric performance. Besides, projection can cause information loss, limiting overall performance~\cite{yang2020graphsim}. Therefore, the performance of these projection-based metrics is not satisfactory under multiple types of distortions.

\par For RR-PCQA, Viola \emph{et al.}~\cite{Irene2020RR2} use the statistical information of the geometry, color and normal vector to evaluate the point cloud quality. Liu \emph{et al.}~\cite{Liu2021RR} build the connection between the quality and compression parameters (e.g., quantification step) of point clouds, which can be used to guide point cloud compression strategy with certain rate constraints.

\par The point cloud quality metrics surveyed above are FR and RR metrics, which means that the input of all these metrics requires the whole reference samples or some of its features. However, in many practical scenarios, obtaining the reference is difficult. For example, some point clouds are captured in the wild, these samples do not have high-quality references naturally. Thus, NR metrics deserve serious treatment. Tao \emph{et al.}~\cite{Tao2021PMBVQA} propose a NR-PCQA metric based on point cloud projection and multi-scale feature fusion. Liu \emph{et al.}~\cite{Liu2021PQANet} propose to predict the quality scores by utilizing the distortion classification information. Yang \emph{et al.}~\cite{Yang2021ITPCQA} propose a NR-PCQA framework by leveraging the rich subjective scores of natural images through the domain adaptation.

\par The existing NR-PCQA metrics are all projection-based, which will introduce information loss~\cite{yang2020graphsim}. In this work, we propose a NR-PCQA metric operating directly on 3D point clouds.

\subsection{3D Feature Description}

\par The first step of learning-based PCQA is to extract the representative features. Early work in 3D applications uses the hand-crafted feature descriptors to discriminate the local geometry characteristic. Johnson \emph{et al.}~\cite{Johnson1999SPIN} propose to project adjacent points onto the tangent plane to describe the geometry characteristics. Tombari \emph{et al.}~\cite{Tombari2010USC} propose to use covariance matrices of point pairs. Salti \emph{et al.}~\cite{Salti2014SHOT} propose to create a 3D histogram of normal vectors. Rusu \emph{et al.} ~\cite{Rusu2008PFH,Rusu2009FPFH} propose to build an oriented histogram using pairwise geometric properties. Guo \emph{et al.}~\cite{Guo2016review} provide a comprehensive review of such hand-crafted descriptors.

\par The powerful representation ability for learning-based methods has attracted more and more attentions. The current research hotspots have been directed to learning-based 3D feature representation. Zeng \emph{et al.}~\cite{Zeng20163DMatch} propose to learn the 3D patch descriptors by leveraging a Siamese CNN. Khoury \emph{et al.}~\cite{Khoury2017Learning} propose to adopt the multi-layer perceptrons to map the 3D oriented histogram to a low-dimensional feature space. Deng \emph{et al.}~\cite{Deng2018PPFNet,Deng2018PPF} propose to adopt PointNet~\cite{Qi2017Pointnet} for geometric feature description.

\par However, the representative feature needed for NR-PCQA is quite different from other 3D applications such as
3D object classification, detection, and segmentation. It requires whole perception and higher efficiency. All the above work extracts a small patch or a set of key points to solve the visual tasks in a lower-dimensional space. However, NR-PCQA requires both local details and global understanding. Specifically, for subjective perception, local stimuli will be first perceived (such as the gradients), and then global stimuli are augmented (e.g., with structural contours of object/scene). Therefore, the design of a NR-PCQA metric should generate the final scores via taking the whole point cloud into consideration.

\par Exploiting the sparsity of 3D point clouds, Graham \emph{et al.}~\cite{Graham2017Submanifold} propose sub-manifold sparse CNN, and illustrate that sub-manifold sparse convolutions offer reliable performance in terms of sparsity invariance and reduced computing load in point cloud processing tasks~\cite{Graham2018submanifold}. Choy \emph{et al.}~\cite{Choy2019Minkowski} propose the Minkowski Engine, an extension of the sub-manifold sparse network to higher dimensions. In this work, the proposed NR-PCQA metric extracts the hierarchical features based on sub-manifold sparse CNN without adopting the dimensionality reduction techniques. In this way, the above-mentioned limitations can be addressed for NR-PCQA.

\section{LS-PCQA: Large-scale Point Cloud Quality Assessment Dataset}\label{sec:dataset}

\par Designing a learning-based NR-PCQA metric needs a large amount of data to improve its robustness and generalization capability. For images, several popular datasets are available, including LIVE~\cite{LIVE}, CSIQ~\cite{Sheikh2006CSIQ}, TID2008~\cite{Wang2006TID2008}, TID2013~\cite{Ponomarenko2015TID2013}, LIVE MD~\cite{Xue2014LIVEMD}, LIVE Challenge~\cite{Gao2016LIVEChallenge} and Waterloo Exploration~\cite{Ma2017Waterloo} for quality assessment under several distortions; CCT~\cite{Min2017CCT} for cross-content-type quality assessment; LIVE-SJTU A/V-QA~\cite{Min2020AVQA} for audio-visual quality assessment; DHQ~\cite{Min2019DHQ} and SHRQ~\cite{Min2019SHRQ} for dehazing quality assessment; OIQA~\cite{Sui2021OIQA} and CVIQ~\cite{Sun2019CVIQ} for omnidirectional images. Among the quality assessment datasets, the Waterloo Exploration has the largest size and it contains 94,880 distorted pictures.

\par However, for 3D point clouds, only a few datasets of small scale have been built. They include PointXR~\cite{Alexiou2020PointXR}, IRPC~\cite{Javaheri2019IRPC}, SJTU-PCQA~\cite{Yang2020TMM3DTO2D} and WPC~\cite{Su2019WPC,Liu2022WPC}, the largest one among them only contains a few hundred distorted samples. The limit in the amount of data can easily lead to overfitting for the learning-based metrics. Besides, the existing datasets have obvious weaknesses. On one hand, the number of reference point clouds and included distortion types are insufficient. On the other hand, the quality of some reference point clouds is not good enough, which can impact the results of subjective experiments. These drawbacks would greatly affect the development of NR-PCQA metrics.

\par The difficulty of building a large-scale PCQA dataset comes from two aspects. The first is to obtain enough original point clouds. To solve this problem, we collect mesh format data and convert them to point clouds to enlarge the availability of point clouds. The second lies in the annotation of built datasets which requires well-conducted subjective experiments under strict control conditions. Since subjective experiments are time-consuming and expensive, it is quite difficult to collect MOS for a large number of point clouds. To address this problem, the pseudo MOS is adopted here. By establishing the annotation criteria, the distorted samples can be automatically labeled by the computer algorithms (i.e., using the high-performance FR metrics).

\par In this paper, 104 original reference point clouds are selected or crafted from ~\cite{Mpeg1,Mpeg2,JPEG,Mpeg3,Nouri2017Greyc,free3d,sketchfab}. Each reference point cloud is distorted by 31 types of impairments under 7 distortion levels. In total, more than 22,000 distorted point clouds are generated. Table~\ref{comparisonScale} compares the newly built dataset and 4 existing datasets, which clearly shows that our newly established dataset has a larger size and covers more distortion types.

\begin{table}[htbp]
  \centering
  \caption{Comparison of existing PCQA datasets and the newly built dataset}
  \begin{footnotesize}
    \begin{tabular}{l|rrr}
    \hline
    Datasets & \multicolumn{1}{l}{Reference samples} & \multicolumn{1}{l}{Distortions} & \multicolumn{1}{l}{Distorted samples} \\
    \hline
    PointXR & 5     & 2     & 40 \\
    IRPC  & 6     & 3     & 54 \\
    SJTU-PCQA & 10    & 7     & 420 \\
    WPC & 20   & 5     & 740 \\
    \hline
    LS-PCQA & 104   & 31    & 22,568 \\
    \hline
    \end{tabular}%
    \end{footnotesize}
  \label{comparisonScale}%
\end{table}%

\par To annotate the new large-scale dataset, we split the whole dataset into three parts. Part I was labeled via the subjective experiment and is used to screen FR metrics for multiple distortion types to generate the pseudo MOS for part II and part III. Part II is used to evaluate the accuracy of pseudo MOS. Part III is labeled using the pseudo MOS only, which is utilized to increase dataset scale. The details are as follows:
\begin{itemize}
 \item Part I contains 930 distorted samples randomly selected from the whole dataset and they are annotated through the subjective experiment. In the sample selection, we pick 6 original point clouds with 5 distortion levels for each of the 31 considered distortion types. The obtained MOS is used to conduct the selection of best FR-PCQA and nonlinear mapping function to compute the pseudo MOS (see Section~\ref{sec:remaininglabeling}).
 \item Part II contains 310 distorted samples, consisting of 2 original point clouds with 5 distortion levels for each of the 31 distortion types. These samples are also annotated through the subjective experiment. We use both the subjective MOS and pseudo MOS to label Part II, and illustrate the accuracy of pseudo MOS (see Section~\ref{sec:accuracyverification}).
 \item Part III contains the remaining distorted samples, which is labeled by the pseudo MOS only. Considering the effectiveness of pseudo MOS, Part III can be extended to arbitrary size to facilitate the construction of large-scale PCQA dataset.
\end{itemize}

\subsection{Reference Point Clouds}

\par
\begin{figure}[htbp]
	\centering
	\includegraphics[width=0.7\linewidth]{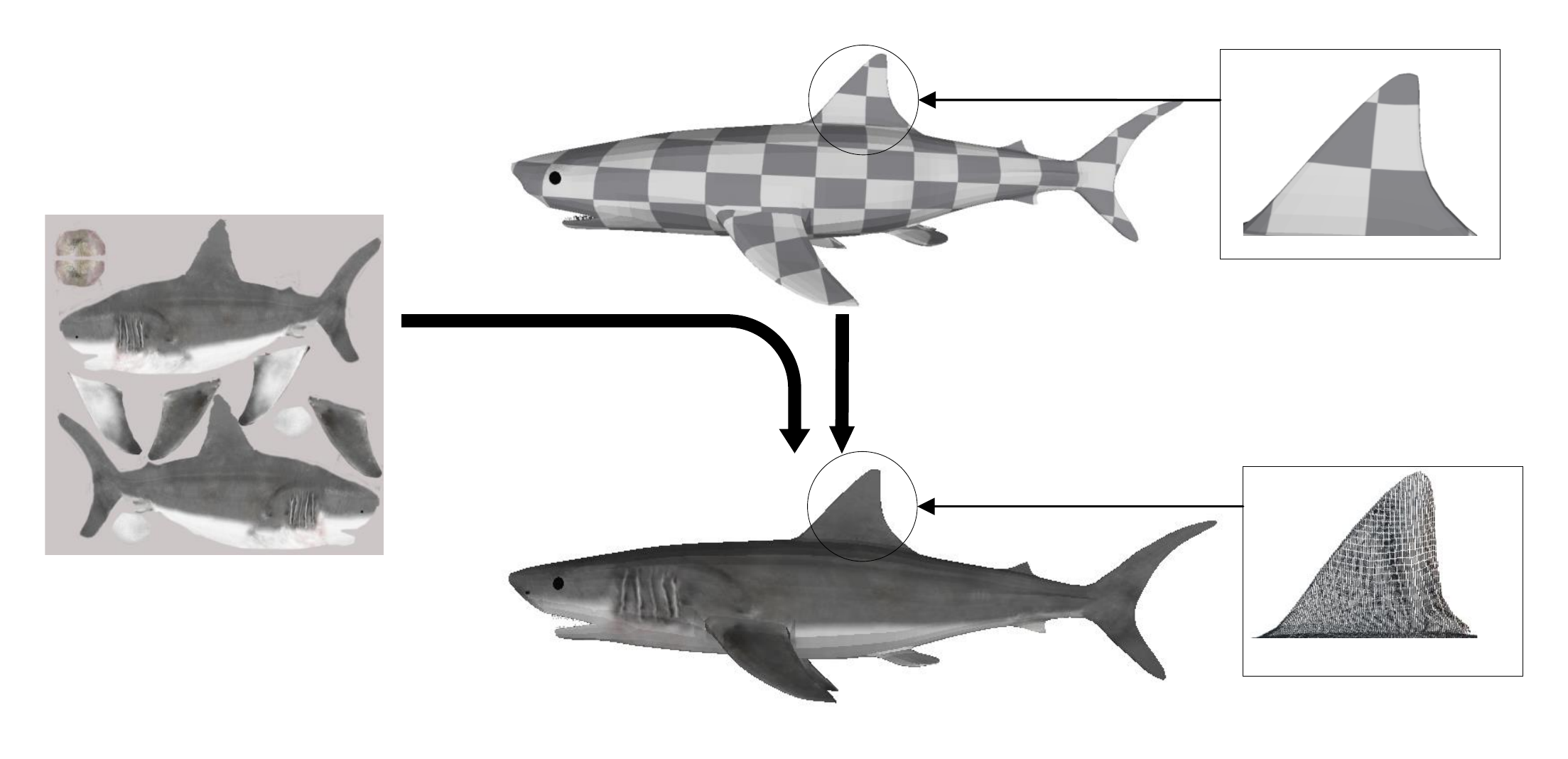}
	\caption{Texture sampling. For generating reference point clouds from mesh format data, we perform random sampling on the surfaces of mesh to generate point clouds. Then the texture material is examined to color the selected points. }
	\label{texturesampling}

\end{figure}

 The reference point clouds in the built dataset come from MPEG and JPEG point cloud datasets~\cite{Mpeg1,Mpeg2,JPEG,Mpeg3}, as well as the 3D mesh data~\cite{Nouri2017Greyc,free3d,sketchfab}. For point clouds from MPEG and JPEG datasets, manual examination is conducted to make sure we obtain point clouds of high quality. We define the point clouds which can score 5 (1 is the lowest and 5 is the highest) in the subjective experiment as high-quality samples. For 3D mesh samples, we apply uniformly distributed random sampling to take sample points from the surfaces of mesh, as shown in Fig. \ref{texturesampling}. Specifically, the surfaces of mesh are randomly sampled to obtain the Cartesian coordinates of the crafted point clouds. For texture, the sampled points are colored by examining the texture material at the same positions.

\par In total, 104 different point clouds, including 28 human models, 48 animal models and 28 inanimate objects, are chosen as the reference point cloud samples. All these reference point clouds are carefully screened to ensure their high quality, all of which score a MOS of 5 in the subjective experiment and don't have holes and other distortions when being presented.

\subsection{Distorted Point Clouds}

\par  Each reference point cloud is degraded by 31 types of impairments under 7 distortion levels, covering a wide range of impairments during point cloud production, compression, transmission and presentation. The distortion types are listed in Table~\ref{SROCC}, and more details are given in the Appendix. In total, 22,568 distorted point clouds are generated.

\subsection{Obtaining the Subjective MOS}

\par To select the best FR-PCQA metric for each distortion type and verify the accuracy of pseudo MOS, we annotate Part I and Part II of the built dataset using the subjective experiment. The double stimulus method is adopted for subjective rating. We strictly follow the steps proposed by ITU-R Recommendation BT. 500~\cite{BT500}. The configuration of the subjective experiment is shown in Table~\ref{Configuration}.

\begin{table}[htbp]
  \centering
  \caption{Setup of the subjective experiment.}
  \begin{footnotesize}
    \begin{tabular}{p{13.055em}|c}
    \hline
    Total number of point clouds & 1,240 \\
    \hline
    Number of scores \newline{}for each point cloud & 16 \\
    \hline
    Viewing time \newline{}for each participant & 0.5h \\
    \hline
    Number of point clouds\newline{} viewed by each participant & 100 \\
    \hline
    Number of participants & 224 \\
    \hline
    \end{tabular}%
    \end{footnotesize}
  \label{Configuration}%
\end{table}%

\par In the subjective experiment, the participants sit in a controlled environment. Specifically, the zoom rate is set as 1:1. The presentation device used in subjective experiments is Dell SE2216H with a 21.5-inch monitor with a resolution of 1920$\times$1080 pixels. Inspired by \cite{Javaheri2019IRPC,Yang2020TMM3DTO2D}, the point-based rendering is adopted with square primitives because of its similarity with the smallest elements of 2D images (pixel), and a size of 2 to ensure no holes in the reference point clouds. The sitting posture of the participants is adjusted to ensure that their eyes are at the same height as the center of the screen. The viewing distance is about three times the height of the rendered point cloud ($\approx$ 0.75 meters). The subjective experiment is conducted indoors, under a normal lighting condition.

\par Only rotation operation is allowed to emulate the free-view navigation in the subjective experiment. This is because the distance to the 3D object will influence the subjective perception significantly. Even for a point cloud with good quality, it will show some holes if we zoom in too much. For participants who are lack of enough prior knowledge on point clouds, this phenomenon will bias their judgment. Therefore, we use the function ‘zoom rate’ provided by CloudCompare to fix the viewing distance, which maintains the consistency of viewing experience across participants.

\par Each pair of point clouds is presented in temporal sequence with the reference always being the first and takes about 20 seconds for each participant to examine, leaving the next 5 seconds for the rating before the next pair is shown. The given scores are in the range of $[1,5]$ which correspond to five quality levels shown in Table~\ref{Score}.

\begin{table}[htbp]
  \centering
  \caption{Five-grade impairment scale.}
  \begin{footnotesize}
    \begin{tabular}{c|l}
    \hline
    MOS     & Meaning \\
    \hline
    5     & Almost no distortion is perceived \\
    \hline
    4     & Distortion can be perceived but don't hinder the viewing \\
    \hline
    3     & Distortion slightly obstructs the viewing \\
    \hline
    2     & Distortion definitely obstructs the viewing \\
    \hline
    1     & Distortion seriously hinders the viewing \\
    \hline
    \end{tabular}%
    \end{footnotesize}
  \label{Score}%
\end{table}%

\par The screening method described in BT. 500~\cite{BT500} is applied to remove the outliers whose scores are inconsistent with others. Specifically, the $\beta2$ test (by calculating the kurtosis coefficient of the function) is adopted to ascertain whether a subject needs to be removed or not. The scores from a subject are rejected when $\beta2$ is not between 2 and 4. As a result, 14 participants are identified as outliers and removed, which may derive from that they were not serious about the experiment pre-training and subjective experiment. The scores from the remaining 210 participants are kept for the following analysis. At least 16 reliable subjective scores are collected for each distorted point cloud after outlier removal.

\par
\begin{figure*}[htbp]
	\centering
	\includegraphics[width=1\linewidth]{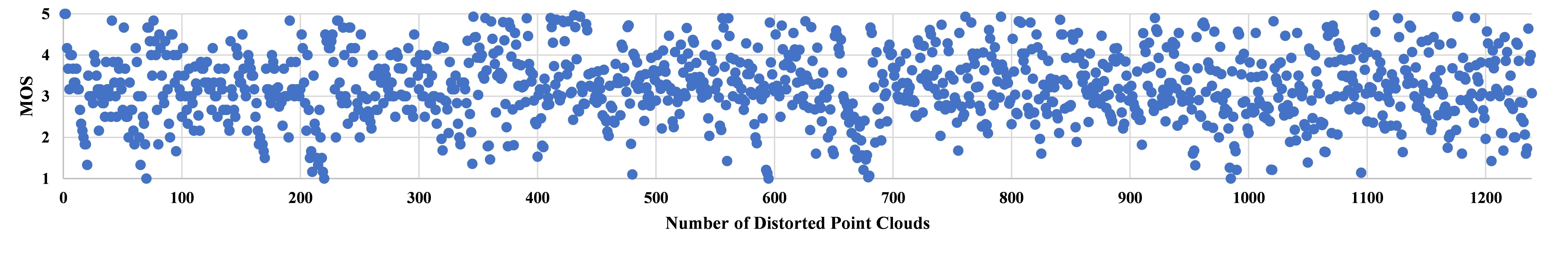}
	\caption{MOS for point clouds in the proposed LS-PCQA.}
	\label{MOS1}

	\centering
	\includegraphics[width=1\linewidth]{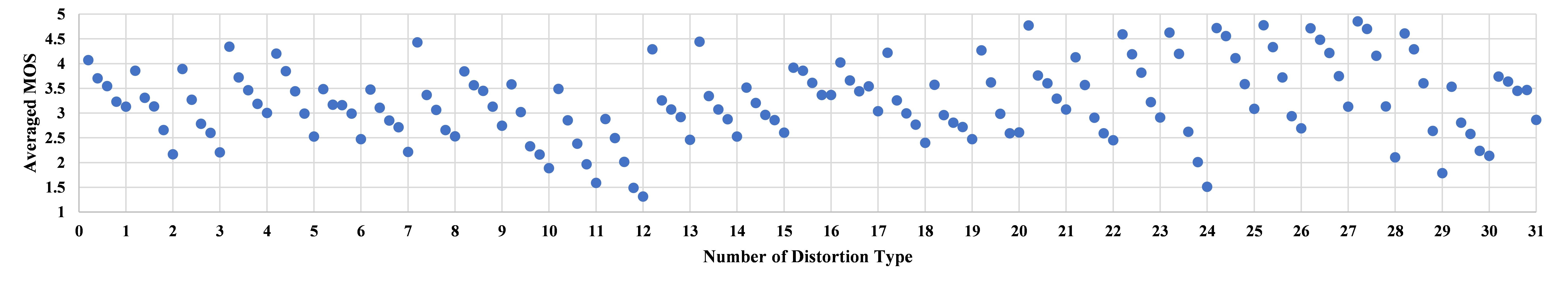}
	\caption{MOS averaged for different distortion levels under each distortion type. The distortion types are indexed in the same order as in Table~\ref{SROCC}.}
	\label{MOS2}

\end{figure*}

 We present the MOS distribution in Fig. \ref{MOS1} and Fig. \ref{MOS2} to demonstrate the validity of subjective data. Fig. \ref{MOS1} shows the MOS distribution for distorted point clouds. It can be seen from Fig. \ref{MOS1} that the subjective scores are spread across various MOS levels. Fig. \ref{MOS2} shows the average MOS for different distortion levels under each distortion type, which demonstrates the correlation between the average MOS and distortion levels.

\subsection{Generation of Pseudo MOS} \label{sec:remaininglabeling}

\subsubsection{Selection of the Best FR-PCQA for Each Distortion Type}

\par FR metrics are used to annotate the point clouds in the built dataset. In order to select the best FR metric for different distortion types, 9 state-of-the-art (SOTA) objective FR-PCQA metrics are examined, including MSE-PSNR-P2point (M-p2po)~\cite{cignoni1998metro,MPEGSoft}, MSE-PSNR-P2plane (M-p2pl)~\cite{Mekuria2016Evaluation,MPEGSoft}, Hausdorff-PSNR-P2point (H-p2po)~\cite{cignoni1998metro,MPEGSoft}, Hausdorff-PSNR-P2plane (H-p2pl)~\cite{Mekuria2016Evaluation,MPEGSoft}, PSNRyuv~\cite{MPEGSoft}, Hausdorff-PSNRyuv (H-PSNRyuv)~\cite{MPEGSoft}, PCQM~\cite{meynet2020pcmd}, GraphSIM~\cite{yang2020graphsim} and MPED~\cite{yang2021MPED}. The Pearson linear correlation coefficient (PLCC) and SROCC are usually used to quantify the performance of the quality assessment metrics.

\par PLCC measures the linear correlation between MOS and predicted quality scores via
\begin{align}
PLCC\left( {{p_i},{{\hat p}_i}} \right){\rm{ }} = \frac{{\sum\limits_i {({p_i} - {p_m})({{\hat p}_i} - {{\hat p}_m})} }}{{\sqrt {\sum\limits_i {{{({p_i} - {p_m})}^2}} } \sqrt {\sum\limits_i {{{({{\hat p}_i} - {{\hat p}_m})}^2}} } }}
\end{align}
where $p_i$ is the true MOS, ${\hat p}_i$ is the predicted quality score, and $p_m$ and ${\hat p}_m$ are their arithmetic mean.

\par SROCC assesses the monotony between MOS and predicted quality scores. It is defined as
\begin{align}
SROCC\left( {{p_i},{{\hat p}_i}} \right){\rm{ }} = 1 - \frac{{6\sum\limits_{i = 1}^L {{{({m_i} - {n_i})}^2}} }}{{L({L^2} - 1)}}
\end{align}
where $L$ is the number of distorted point clouds, $m_i$ is the rank of $p_i$ in the MOS, and $n_i$ is the rank of ${\hat p}_i$ in the predicted quality scores. SROCC can be reformulated as
\begin{align}
SROCC\left( {{p_i},{{\hat p}_i}} \right){\rm{ }} = {\rm{ }}PLCC\left( {R\left( {{p_i}} \right),R\left( {{{\hat p}_i}} \right)} \right)
\end{align}
where $R$ denotes the ranking operation.

\par SROCC is considered to be the best nonlinear correlation indicator, because SROCC is only concerned with the order of elements in a sequence. Therefore, even if ${p_i}$ or ${{\hat p}_i}$ is affected by any monotone nonlinear transformation (such as logarithmic transformation and exponential transformation), SROCC will not be affected because the order of elements is not changed. Therefore, we use SROCC as index to select the best FR metrics. Table~\ref{SROCC} lists the SROCC of these FR-PCQA metrics for each distortion type based on the subjective MOS of Part I, and the best results are highlighted in bold.

\begin{table*}[t]
  \centering
  \caption{SROCC of FR-PCQAs for each distortion type.}
  \resizebox{\textwidth}{!}{
    \begin{tabular}{c|p{8em}|c|c|c|c|c|c|c|c|c}
    \hline
    No    & Distortion Type & M-p2po & M-p2pl & H-p2po & H-p2pl & PSNRyuv & H-PSNRyuv & PCQM  & GraphSIM & MPED \\
    \hline
    \#1   & ColorNoise & -     & -     & -     & -     & 0.833986 & \textbf{0.847526} & 0.815255 & 0.686812 & 0.786489  \\
    \hline
    \#2   & ColorQuantization\newline{}Dither & -     & -     & -     & -     & \textbf{0.825952} & 0.749026 & 0.769419 & 0.516804 & 0.638994  \\
    \hline
    \#3   & ContrastDistortion & -     & -     & -     & -     & 0.689448 & \textbf{0.765722} & 0.743989 & 0.678540 & 0.374221  \\
    \hline
    \#4   & CorrelatedGaussian\newline{}Noise & -     & -     & -     & -     & \textbf{0.939239} & 0.915869 & 0.850657 & 0.589139 & 0.665702  \\
    \hline
    \#5   & DownSample & \textbf{0.881299} & 0.626544 & 0.841531 & 0.811972 & 0.014907 & -0.016913 & 0.524864 & 0.842585 & 0.857492  \\
    \hline
    \#6   & GammaNoise & -     & -     & -     & -     & \textbf{0.749555} & 0.682405 & 0.707703 & 0.588379 & 0.733081  \\
    \hline
    \#7   & GaussianNoise & -     & -     & -     & -     & 0.919226 & 0.903183 & \textbf{0.950823} & 0.769025 & 0.342164  \\
    \hline
    \#8   & GaussianShifting & 0.741433 & 0.718959 & 0.829773 & \textbf{0.834223} & 0.755229 & 0.235674 & 0.816199 & 0.742768 & 0.598353  \\
    \hline
    \#9   & HighFrequencyNoise & -     & -     & -     & -     & 0.836319 & 0.803160 & \textbf{0.915100} & 0.762880 & 0.800712  \\
    \hline
    \#10  & LocalMissing & 0.536227 & 0.497941 & 0.036992 & 0.302997 & 0.689371 & 0.364701 & 0.770840 & \textbf{0.871007} & 0.565895  \\
    \hline
    \#11  & LocalOffset & \textbf{0.937340} & 0.934669 & 0.770490 & 0.770738 & 0.667112 & 0.019499 & 0.851642 & 0.906177 & 0.897719  \\
    \hline
    \#12  & LocalRotation & 0.819668 & 0.712649 & \textbf{0.831126} & 0.734499 & 0.327066 & 0.004901 & 0.657248 & 0.723854 & 0.742183  \\
    \hline
    \#13  & LumaNoise & -     & -     & -     & -     & 0.772672 & \textbf{0.855360} & 0.748414 & 0.817848 & 0.715923  \\
    \hline
    \#14  & MeanShift & -     & -     & -     & -     & 0.422259 & \textbf{0.818676} & 0.614734 & 0.706210 & 0.727131  \\
    \hline
    \#15  & Multiplicative\newline{}GaussianNoise & -     & -     & -     & -     & 0.751224 & 0.563910 & \textbf{0.754339} & 0.648420 & 0.687361  \\
    \hline
    \#16  & PoissonNoise & -     & -     & -     & -     & \textbf{0.682406} & 0.680179 & 0.662584 & 0.421827 & 0.366815  \\
    \hline
    \#17  & QuantizationNoise & -     & -     & -     & -     & 0.780537 & 0.398172 & \textbf{0.848013} & 0.617971 & 0.709498  \\
    \hline
    \#18  & RayleighNoise & -     & -     & -     & -     & \textbf{0.893835} & 0.804362 & 0.837748 & 0.706877 & 0.732918  \\
    \hline
    \#19  & SaltpepperNoise & -     & -     & -     & -     & 0.394971 & 0.044449 & 0.637517 & 0.560080 & \textbf{0.698709 } \\
    \hline
    \#20  & SaturationDistortion & -     & -     & -     & -     & 0.739290 & 0.791143 & \textbf{0.850562} & 0.703016 & 0.597530  \\
    \hline
    \#21  & UniformNoise & -     & -     & -     & -     & \textbf{0.898209} & 0.871511 & 0.685727 & 0.645456 & 0.714651  \\
    \hline
    \#22  & UniformShifting & 0.851675 & 0.851675 & 0.857461 & 0.849672 & 0.796929 & 0.404542 & 0.638923 & 0.869701 & \textbf{0.890175 } \\
    \hline
    \#23  & VPCC\_lossy-geom-lossy-attrs & 0.753449 & 0.753004 & 0.640854 & 0.632844 & 0.372052 & -0.005563 & 0.795505 & \textbf{0.808189} & 0.731420  \\
    \hline
    \#24  & AVS\_limitlossyG-lossyA & 0.934698 & \textbf{0.945823} & 0.939598 & 0.929384 & 0.882412 & 0.337524 & -     & 0.743907 & 0.937590  \\
    \hline
    \#25  & AVS\_losslessG-limitlossyA & -     & -     & -     & -     & 0.837784 & \textbf{0.885227} & -     & 0.877170 & 0.815754  \\
    \hline
    \#26  & AVS\_losslessG-lossyA & -     & -     & -     & -     & 0.916352 & \textbf{0.936927} & -     & 0.795773 & 0.870078  \\
    \hline
    \#27  & GPCC\_lossless-geom-lossy-attrs & -     & -     & -     & -     & 0.567931 & 0.734839 & \textbf{0.862134} & 0.664515 & 0.517192  \\
    \hline
    \#38  & GPCC\_lossless-geom-nearlossless-attrs & -     & -     & -     & -     & 0.878282 & 0.932864 & \textbf{0.937695} & 0.899421 & 0.817312  \\
    \hline
    \#29  & GPCC\_lossy-geom-lossy-attrs & 0.955370 & 0.925543 & \textbf{0.960451} & 0.946853 & 0.730774 & 0.197663 & 0.844964 & 0.867446 & 0.903283  \\
    \hline
    \#30  & Octree & 0.779039 & 0.788162 & \textbf{0.819856} & 0.752114 & 0.523810 & 0.108589 & 0.676235 & 0.757454 & 0.710948  \\
    \hline
    \#31  & Possion\newline{}Reconstruction & \textbf{0.847028} & 0.835449 & 0.812514 & 0.811401 & 0.213093 & -0.137609 & 0.720553 & 0.647518 & 0.719662  \\
    \hline
    \end{tabular}%
    }
  \label{SROCC}%
\end{table*}%

\par Table~\ref{SROCC} indicates that adopting a single quality assessment metric to label the whole dataset will be insufficient and inaccurate. Each quality assessment metric has its own limitations. Although some FR metrics achieve the best results for certain distortion types, they may have poor performance or even cannot respond to some other distortion types. For example, p2point has achieved the best performance for some geometrical distortions, but it is insensitive to photometric attribute distortions. PCQM achieves a SROCC of 0.950823 for Gaussian noise distortion, but for down-sampling distortion, its SROCC is only 0.524864. Thus, we keep the best one among these FR-PCQA metrics for each distortion type to annotate the distorted point clouds.

\subsubsection{Nonlinear Mapping}

\par The range of each FR-PCQA metric is different from each other. To label the distorted point clouds, a nonlinear regression function is adopted to re-scale the results to cast them under the same range of [1,5]. In this process, the nonlinear regression function should not change the monotonicity of the scores from the FR metrics, but the overall monotonicity may be different for different nonlinear regression functions due to the splicing of the quality scores. The commonly used nonlinear regression functions include four-parameter logistic regression (Logistic-4)~\cite{2000logistic4}, five-parameter logistic regression (Logistic-5)~\cite{Sheikh2006CSIQ}, and four-parameter polynomial regression (Cubic-4)~\cite{2012cubic4,2010cubic4}.

\par Logistic-4 applies
\begin{align}
Q = \frac{{{\beta _1} - {\beta _2}}}{{1 + {e^{ - \frac{{Q_s - {\beta _3}}}{{\left| {{\beta _4}} \right|}}}}}} + {\beta _2}
\end{align}
where $Q$ is the normalized score, $Q_s$ is the quality score predicted by the best quality assessment metric and ${\beta _1,\beta _2,\beta _3,\beta _4}$ are the fitting parameters.

\par Logistic-5 achieves normalization via
\begin{align}
Q = {\beta _1}\left( {\frac{1}{2} - \frac{1}{{1 + {e^{{\beta _2}({Q_s} - {\beta _3})}}}}} \right) + {\beta _4}{Q_s} + {\beta _5}
\end{align}
where ${\beta _1,\beta _2,\beta _3,\beta _4,\beta _5}$ are the parameters to be determined.

\par Cubic-4 uses the following for normalization
\begin{align}
Q = a{Q_s}^3 + b{Q_s}^2 + c{Q_s} + d
\end{align}
where ${a,b,c,d}$ are the model parameters.

\par In order to select the most appropriate mapping function, a validation experiment is conducted using Part I of the newly built dataset. The three nonlinear regression functions in consideration are adopted for MOS normalization to project the best-predicting FR metric for each distortion type to the range of subjective MOS (i.e., 1-5). Then, SROCCs between the normalized scores and subjective MOS are computed for each nonlinear regression function. The results are listed in Table~\ref{nonlinear}, with the best results highlighted in bold.

\begin{table}[htbp]
  \centering
  \caption{Comparison of regression results for three nonlinear regression functions in consideration.}
  \begin{footnotesize}
    \begin{tabular}{l|rrr}
    \hline
          & \multicolumn{1}{l}{Logistic-4} & \multicolumn{1}{l}{Logistic-5} & \multicolumn{1}{l}{Cubic-4} \\
    \hline
    SROCC & 0.857299 & \textbf{0.902697} & 0.895769 \\
    PLCC  & 0.862648 & \textbf{0.910713} & 0.904457 \\
    \hline
    \end{tabular}%
    \end{footnotesize}
  \label{nonlinear}%
\end{table}%

\par It can be seen from Table~\ref{nonlinear} that Logistic-5 offers the best results. Thus, Logistic-5 is adopted as the nonlinear normalization function for MOS mapping in this work. The normalized pseudo MOS is used as the labels for the built large-scale dataset.

\subsection{Accuracy Analysis of Pseudo MOS} \label{sec:accuracyverification}

\par We verify the reliability of the pseudo MOS via conducting the validation experiment using samples from Part I and Part II of the partitioned dataset. Note that Part II of the dataset is not used in the selection of FR-PCQA metrics to generate the pseudo MOS. The SROCCs between the subjective MOS and generated pseudo MOS in Part I and Part II are computed and summarized in Table~\ref{pseudoverification}.

\begin{table}[htbp]
  \centering
  \caption{Correlation between the pseudo and subjective MOS over Part I and Part II of the new dataset. Both of them present high consistency.}
  \begin{footnotesize}
    \begin{tabular}{l|cc}
    \hline
          & Part I & Part II \\
    \hline
    SROCC & 0.902697 & 0.878517 \\
    PLCC  & 0.910713 & 0.871917 \\
    \hline
    \end{tabular}%
    \end{footnotesize}
  \label{pseudoverification}%
\end{table}%

\par We can see from Table~\ref{pseudoverification} that the SROCCs between the subjective and pseudo MOS of Part I and Part II are 0.902697 and 0.878517, which justifies the accuracy of the computed pseudo MOS and validity of the proposed annotation method.

\par
\begin{figure}[htbp]
\centering
  \subfigure[]{\includegraphics[width=0.49\linewidth]{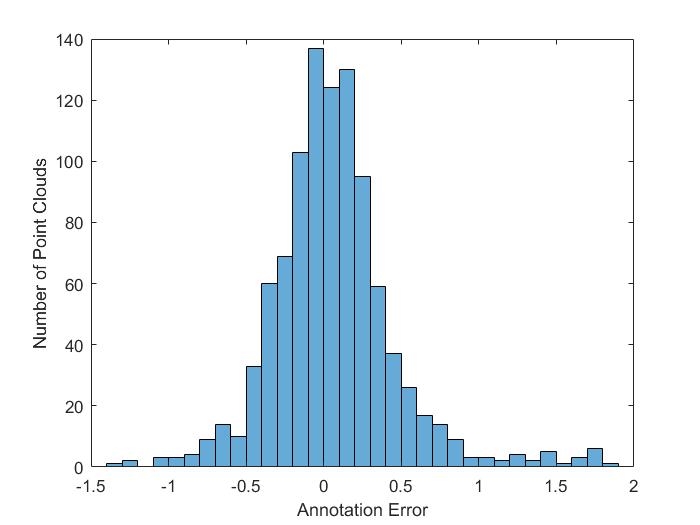}\label{sfig:predictionErrorPartI}}
  \subfigure[]{\includegraphics[width=0.49\linewidth]{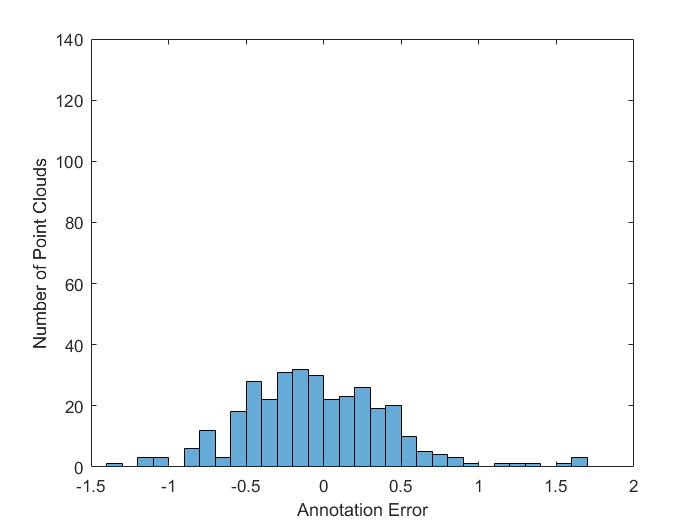}\label{sfig:predictionErrorPartII}}
  \centering
\caption{{\bf Histograms of annotation errors.} (a) for Part I; (b) for Part II.}
  \label{histogramsMOSandPseusoMOS}
\end{figure}

 Statistical analysis of annotation errors for samples in Part I and Part II is also conducted. The annotation error is defined as $(MOS - pseudo \, MOS)$. The histograms of annotation errors for Part I and Part II are shown in Fig \ref{histogramsMOSandPseusoMOS}. The mean, standard deviation and $95\%$ quantile of annotation errors for Part I and Part II are summarized in Table~\ref{StatisticalAnalysis}. Fig. \ref{histogramsMOSandPseusoMOS} shows that most annotation errors have small magnitudes, even for the samples in Part II that are not used in the selection of FR metrics. These results corroborate that the pseudo MOS can be considered accurate approximation of the costly subjective MOS.

\begin{table}[htbp]
  \centering
  \caption{Mean, standard deviation and $95\%$ quantile of the annotation errors.}
  \begin{footnotesize}
    \begin{tabular}{l|r|r}
    \hline
          & \multicolumn{1}{l|}{Part I} & \multicolumn{1}{l}{Part II} \\
    \hline
    Mean  & 0.0653 & -0.0435 \\
    \hline
    Standard deviation & 0.4057 & 0.4804 \\
    \hline
    95\% quantile & 0.7209 & 0.6752 \\
    \hline
    \end{tabular}%
    \end{footnotesize}
  \label{StatisticalAnalysis}%
\end{table}%

\par To gain more insights, we present the annotation error statistics as a function of the distortion levels in Table~\ref{StatisticalWithDisLevel}. We can see that the pseudo MOS exhibits improved accuracy under more severe distortions. This is in fact expected, as the HVS would be more sensitive to obvious distortions, which can be better captured by FR metrics.

\begin{table}[htbp]
  \centering
  \caption{Mean, standard deviation and $95\%$ quantile of the annotation errors under different distortion levels.}
  \begin{footnotesize}
    \begin{tabular}{l|c|c|c|c|c}
    \hline
    Distortion level & Slight & Low   & Medium & High  & Serious \\
    \hline
    Mean  & 0.1433 & 0.102 & 0.0627 & 0.0158 & 0.0029 \\
    \hline
    Standard deviation   & 0.4315 & 0.4322 & 0.4070 & 0.3522 & 0.3870 \\
    \hline
    95\% quantile & 1.0058 & 0.8068 & 0.6877 & 0.5841 & 0.6533 \\
    \hline
    \end{tabular}%
    \end{footnotesize}
  \label{StatisticalWithDisLevel}%
\end{table}%

\par
\begin{figure*}[htbp]
	\centering
	\includegraphics[width=1\linewidth]{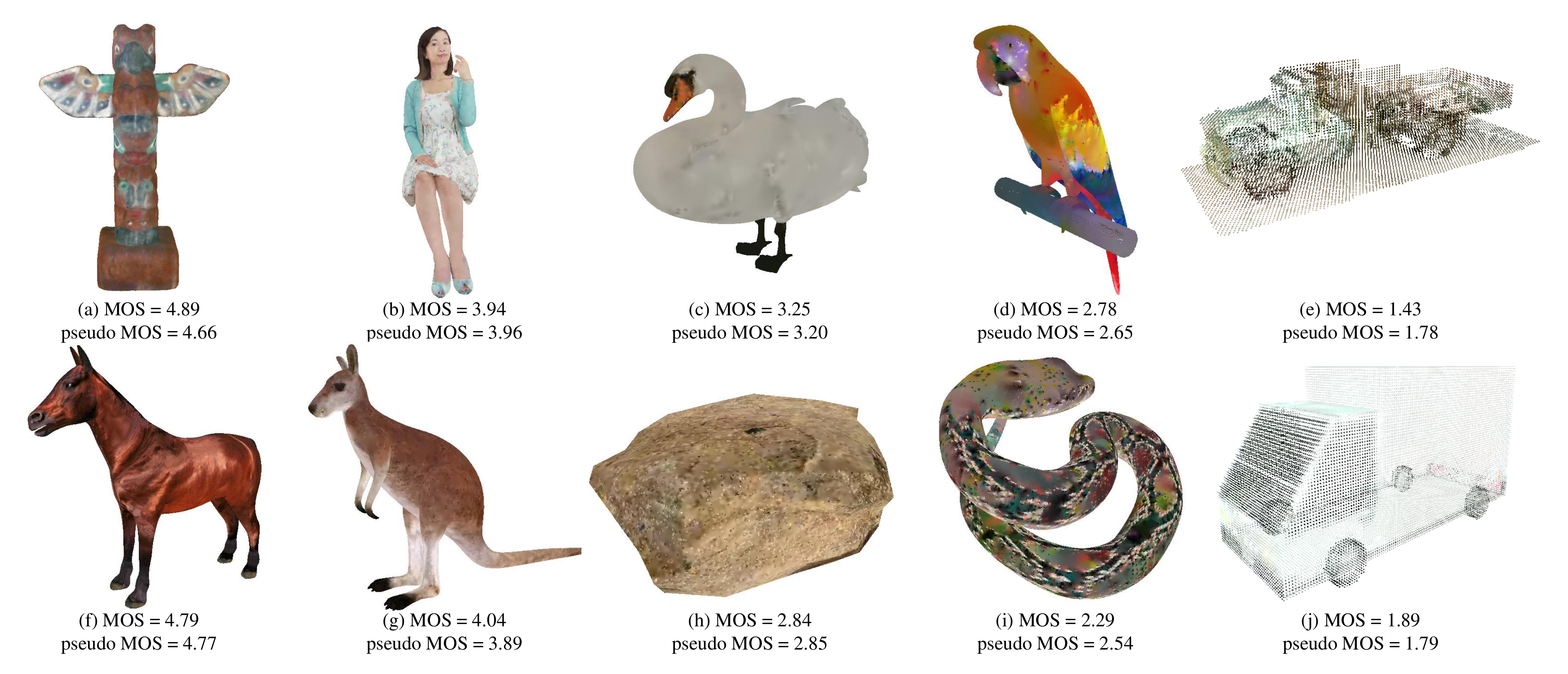}
	\caption{Selected point clouds with their subjective MOS and pseudo MOS. The associated subjective MOS and pseudo MOS are consistent.}
	\label{pseudomosandmos}

\end{figure*}

 Finally, we show some examples of distorted point clouds with subjective MOS and pseudo MOS in the newly built dataset in Fig~\ref{pseudomosandmos}. It can be observed that the subjective MOS and pseudo MOS are quite close to each other in these examples.

\section{ResSCNN: A Sparse CNN Based Metric for NR-PCQA}\label{sec:framework}

\par An end-to-end learning-based NR-PCQA metric ResSCNN is proposed in this section.

\subsection{Network Architecture}

\par
\begin{figure}[htbp]
	\centering
	\includegraphics[width=0.7\linewidth]{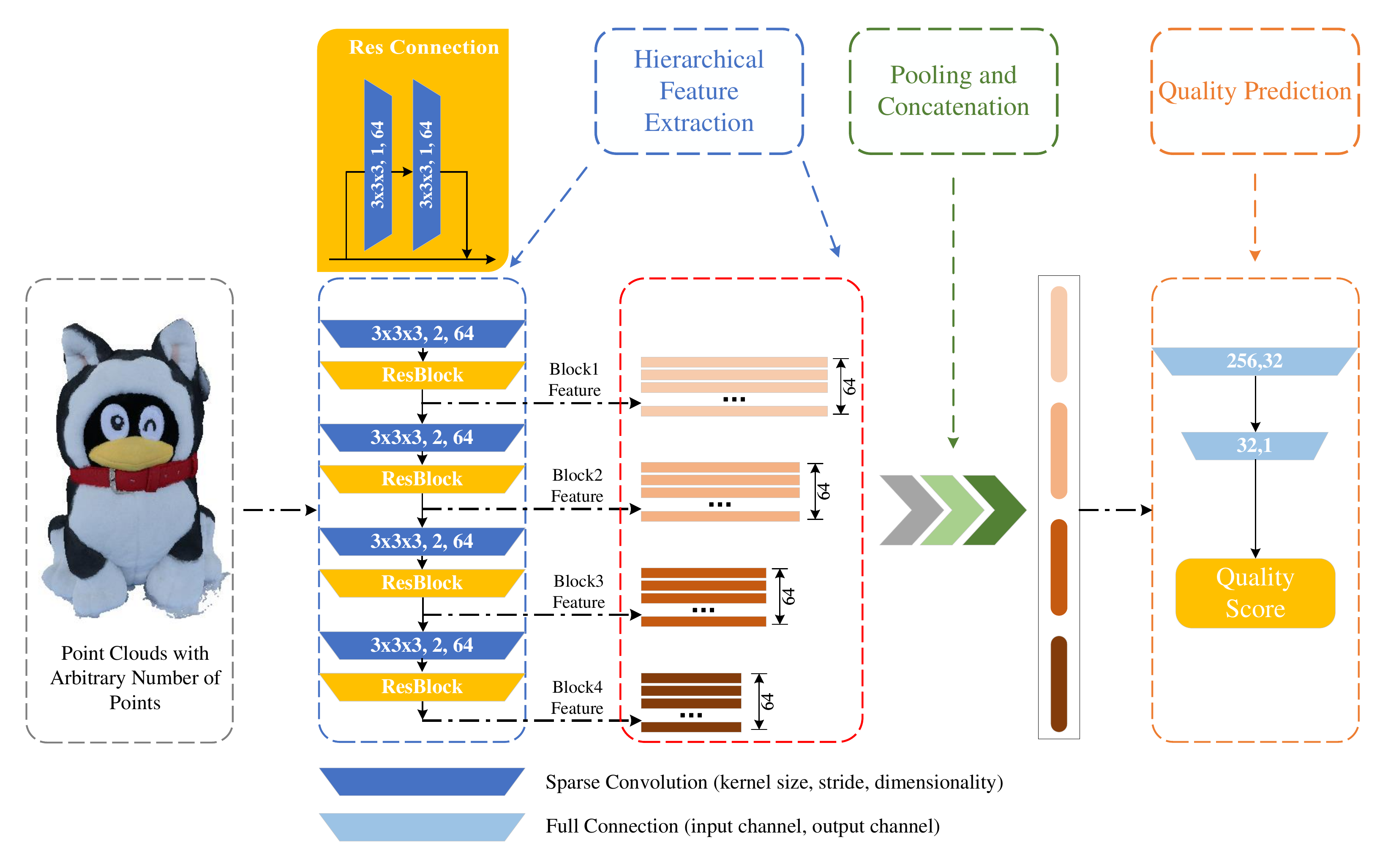}
	\caption{Architecture of ResSCNN. It consists of the following substeps: (a) The complete 3D point clouds are fed into the network to avoid extra distortions introduced by the dimensionality reduction techniques. (b) The hierarchical features are extracted from a designed sparse CNN to avoid the massive increase in the elements of feature map after the conventional dense convolution. (c) The extracted features are pooled and concatenated to generate the feature vector with fixed shape. (d) The quality prediction module achieves the prediction of the point cloud quality.}
	\label{framework}

\end{figure}

 As shown in Fig. \ref{framework}, the proposed ResSCNN consists of three modules: a hierarchical feature extraction module $W^{f}$, a pooling and concatenation module $\Phi$ and a quality prediction module $W^{r}$.
$W^{f}$ takes point clouds with arbitrary number of points as input and uses a stack of sparse convolutional layers and residual blocks to extract the hierarchical features. $\Phi$ applies the global pooling and concatenation operation to generate the feature vectors with consistent shape. $W^{r}$ uses a concatenation of fully connected layers to map the feature vectors to the predicted quality scores.

\subsection{Hierarchical Feature Extraction}

\par The hierarchical feature extraction module is composed of four blocks. Each block consists of three sparse convolutional layers, in which the second and third sparse layers are connected to the residual pattern. The input of feature extraction module is the sparse tensor of the point cloud with arbitrary number of points. Mathematically, a sparse tensor for point cloud $P \in {R^{N \times 6}}$ is represented as a set of coordinates $C$ and associated features $F$:

\begin{align}
C = \left[ {\begin{array}{*{20}{c}}
{{x_1}}&{{y_1}}&{{z_1}}&{{b_1}}\\
 \vdots & \vdots & \vdots & \vdots \\
{{x_N}}&{{y_N}}&{{z_N}}&{{b_N}}
\end{array}} \right],F = \left[ {\begin{array}{*{20}{c}}
{f_1^T}\\
 \vdots \\
{f_N^T}
\end{array}} \right]
\end{align}
where $C$ indicates the geometry attributes of point cloud and ${x_i},{y_i},{z_i} \in Z$ are the 3D coordinates of point $i$. ${b_i}$ is the occupation index to distinguish points occupying the same coordinates. $F$ in the input sparse tensor has 3 feature dimensionalities referring to the color attributes of point cloud and ${f_i} \in {N^{3\times 1 }},{f_i} \in [0,255]$ is the R, G, B attributes associated with the $i$-th point.

\par The $3DConv$ in the proposed network refers to the sparse convolution, which can be denoted generically as
\begin{align}
f_u^{out} = \sum\limits_{i \in {N^D}(u,{C^{in}})} {{W_i}f_{u + i}^{in}~\text{for}~u \in {C^{out}}}
\end{align}
where ${N^D}(u,{C^{in}}) = \{ i|u + i \in {C^{in}},i \in {N^D}\} $ defines the shape of a 3D convolutional kernel, covering the set of offsets from the current center, $u$, such that $u \in {C^{in}}$. $f_u^{in}$ and $f_u^{out}$ denote the feature vector at the coordinate $u$. ${C^{in}}$ and ${C^{out}}$ are the predefined input and output coordinates of sparse tensors. ${W_i}$ denotes the kernel value at offset $i$.
\par In this work, we set $C^{in}=C^{out}=C$ and set $N^D$ as the list of offsets in D-dimensional hypercube centered at the origin to achieve the sub-manifold sparse convolution~\cite{Graham2017Submanifold}.

\par The sparse convolution, instead of the dense convolution, is adopted to extract the features of the input point clouds. The sparse convolution will not reduce the sparsity of point clouds. On the contrary, due to the density heterogeneity, the conventional dense convolution will result in a massive increase in the elements of feature map for point clouds.

\par Each sparse convolutional layer in Fig. \ref{framework} is characterized by the kernel size, stride and dimensionality, and is followed by batch normalization before being fed to the nonlinearity activation function (ReLU).

\subsection{Pooling and Concatenation}

\par The branches of pooling and concatenation module come from the output of each block in the hierarchical feature extraction module. To deal with the output of each block with different shapes, the features extracted from the sparse CNN are globally pooled into $64\times 1$ feature vectors. Then the obtained four $64\times 1$ feature vectors are concatenated to form the $256\times 1$ representative hierarchical feature vector.

\subsection{Quality Prediction}

\par The quality prediction module is composed of two fully connected layers, denoted as FC-1 and FC-2, to map the hierarchical feature vector to the predicted quality score. The number of input channels of FC-1 is 256, which is equal to the length of the hierarchical features. The number of output channels of FC-1 and the number of input channels of FC-2 are 32. The output of the quality prediction module is a predicted quality score of 1 channel with the same scale of training labels.
\par In the pooling and concatenation module, global pooling is applied to obtain the feature vector from different depths of the sparse CNN, which can be denoted as
\begin{align}
{S_i} = {\Phi _i}(X;{W^f}),i = 1, \cdots ,4
\end{align}
where ${S_i} \in {R^{64 \times 1}}$ represents the normalized feature vector from different depths of the hierarchical feature extraction module, $W^{f}$ denotes the parameters of the hierarchical feature extraction module, and the operation $\Phi _i$ denotes the global pooling applied to the original deep convolutional features with irregular shapes. The hierarchical feature vectors obtained from the pooling and concatenation module are
\begin{align}
S = {S_1} \oplus {S_2} \oplus {S_3} \oplus {S_4}
\end{align}
where $\oplus$ denotes the concatenation operation. The final predicted quality scores are found via
\begin{align}
Q = F(S;{W^r})
\end{align}
where $Q$ is the predicted quality score, $F$ represents the fully connected layers, and $W^r$ denotes the parameters of the quality prediction module.

\subsection{Loss Function and Training Strategy}

\par $L_1$ loss, $L_1(x)=|x|$, will fluctuate around the stable value, and converging to achieve higher accuracy is difficult. The gradient of $L_2$ loss $L_2(x)=x^2$ with respect to the predicted value at the beginning of training is large and the training is unstable. Since smooth $L_1$ loss combines the advantages of $L_1$ loss and $L_2$ loss, and avoids their disadvantages, smooth $L_1$ loss is adopted to improve the robustness of the network:
\begin{align}
Smooth{L_1}(x) = \left\{ {\begin{array}{*{20}{c}}
{0.5{x^2},~\text{if}\left| x \right| < 1}\\
{\left| x \right| - 0.5,~\text{otherwise}}
\end{array}} \right.
\end{align}
where $x = q - \overline q $, $q$ is the predicted quality score and $\overline q$ is the ground truth. The gradient of loss function with respect to $x$ is formulated as
\begin{align}
\frac{{\partial smooth{L_1}(x)}}{{\partial x}} = \left\{ {\begin{array}{*{20}{c}}
{x,~\text{if}\left| x \right| < 1}\\
{ \pm 1,~\text{otherwise}}
\end{array}} \right.
\end{align}
where we can see that when $x = q - \overline q $ is small, the gradient with respect to $x$ becomes smaller. When $x$ is large, the upper limit of the gradient with respect to $x$ is 1, to ensure that derivatives are continuous for all degrees.

\par The training of ResSCNN for point clouds consumes much more time than that of training using 2D images. To accelerate the speed of training, Stochastic Gradient Descent (SGD) is adopted with a learning rate $lr = 1e-3$ with an exponential learning rate schedule $\gamma  = 0.99$.

\par As the shapes of input point clouds are different from one another for end-to-end learning, the batch size is set to 1 when training the proposed network. We accumulate several losses and gradients to emulate the process of batch optimization. Besides, the data augmentation is invoked during training with random scaling of $[0.8,1.2]$ and random rotation of $[{0^\circ },{360^\circ })$ to make sure that the proposed network is robust to the transformation of viewpoint.

\section{Experiments}\label{sec:experiment}

\subsection{Experiment Setups}

\subsubsection{Datasets}

\par To evaluate the performance of the proposed ResSCNN, we conduct evaluation experiments using Part III of the proposed LS-PCQA dataset, where the distorted samples are all labeled by pseudo MOS. Besides, the performance of the proposed metric is also evaluated over SJTU-PCQA~\cite{Yang2020TMM3DTO2D} and WPC2.0~\cite{Su2019WPC,Liu2022WPC} datasets.

\textbf{SJTU-PCQA.} SJTU-PCQA includes 10 reference point clouds with 7 different types of distortions under 6 levels, including 4 individual distortions and 3 superimposed distortions. In total, there are 420 distorted point cloud samples.

\textbf{WPC2.0.} WPC2.0 includes 16 reference point cloud samples with V-PCC distortion. 5 geometry quantization steps and 5 texture quantization steps are considered. In total, there are 400 distorted samples.

\subsubsection{Implementation Details}

 \par To compare the proposed ResSCNN with other learning-based NR-PCQA metrics, we split LS-PCQA, SJTU-PCQA and WPC2.0 into training sets and testing sets. The training set and testing set from LS-PCQA contain the distorted samples generated from 100 reference point clouds and 4 reference point clouds respectively to avoid overlapping. For SJTU-PCQA and WPC2.0, we split the reference point clouds of two datasets in a way such that $75\%$ of the samples are used for training while the remaining $25\%$ are for testing.

\par The scale of training labels is not normalized, and the network will automatically learn and predict the quality scores within the scale of training labels. For LS-PCQA dataset, the scale of input and output scores is [1, 5]. For SJTU-PCQA dataset, the scale of input and output scores is [1, 10]. For WPC2.0 dataset, the scale of input and output scores is [0, 100].

\subsection{Prediction Performance}\label{sec:LSexperiment}

\par We use PLCC and SROCC to quantify the performance of the quality assessment metrics. In particular, we compare the performance of the proposed ResSCNN with the existing FR metrics, including MSE-PSNR-P2point (M-p2po)~\cite{cignoni1998metro,MPEGSoft}, MSE-PSNR-P2plane (M-p2pl)~\cite{Mekuria2016Evaluation,MPEGSoft}, Hausdorff-PSNR-P2point (H-p2po)~\cite{cignoni1998metro,MPEGSoft}, Hausdorff-PSNR-P2plane (H-p2pl)~\cite{Mekuria2016Evaluation,MPEGSoft}, PSNRyuv~\cite{MPEGSoft}, Hausdorff-PSNRyuv (H-PSNRyuv)~\cite{MPEGSoft}, PCQM~\cite{meynet2020pcmd}, GraphSIM~\cite{yang2020graphsim} and MPED~\cite{yang2021MPED}, as well as  the existing NR metrics, including PQA-Net~\cite{Liu2021PQANet} and IT-PCQA~\cite{Yang2021ITPCQA}. All the FR-PCQA and NR-PCQA results are obtained using the source code released by the authors. The results are summarized in Table~\ref{comparedwithothers}, where the rows represent the quality assessment metrics and columns give the results of overall SROCC and PLCC.

\begin{table}[htbp]
  \centering
  \caption{Metric performance (PLCC and SROCC) on LS-PCQA, SJTU-PCQA and WPC2.0 datasets.}
  \begin{footnotesize}
    \begin{tabular}{c|l|rr|rr|rr}
    \hline
    \multicolumn{2}{c|}{\multirow{2}[4]{*}{}} & \multicolumn{2}{c|}{LS-PCQA} & \multicolumn{2}{c|}{SJTU-PCQA} & \multicolumn{2}{c}{WPC2.0} \\
\cline{3-8}    \multicolumn{2}{c|}{} & \multicolumn{1}{l}{PLCC} & \multicolumn{1}{l|}{SROCC} & \multicolumn{1}{l}{PLCC} & \multicolumn{1}{l|}{SROCC} & \multicolumn{1}{l}{PLCC} & \multicolumn{1}{l}{SROCC} \\
    \hline
    \multirow{9}[2]{*}{Full-reference} & M-p2po & 0.61&0.30&0.88&0.78&0.61&0.58  \\
        &M-p2pl & 0.59&0.29&0.77&0.63&0.63&0.59  \\
        &H-p2po & 0.52&0.26&0.60&0.58&0.51&0.46  \\
        &H-p2pl & 0.53&0.26&0.65&0.62&0.55&0.48  \\
        &PSNRyuv & 0.63&0.62&0.65&0.62&0.46&0.47  \\
        &H-PSNRyuv & 0.60&0.57&0.43&0.41&0.29&0.23  \\
        &PCQM&0.40&0.54&0.84&0.83&\textbf{0.74} & \textbf{0.75} \\
        &GraphSIM & 0.43&0.46&\textbf{0.91} & \textbf{0.89} & \textbf{0.74} & \textbf{0.75} \\
        &MPED&\textbf{0.70} & \textbf{0.67} &0.37&0.31&0.60&0.59  \\
    \hline
    \multirow{3}[2]{*}{No-reference}
     & PQA-Net &0.56 &0.52 &0.28 &0.23 & - & -  \\
      & IT-PCQA &0.51&0.53&0.58&0.63&0.55&0.54  \\
        &ResSCNN & \textbf{0.60} & \textbf{0.62} & \textbf{0.86} & \textbf{0.81} & \textbf{0.72} & \textbf{0.75} \\
    \hline
    \end{tabular}%
    \end{footnotesize}
  \label{comparedwithothers}%
\end{table}%

\par From Table~\ref{comparedwithothers}, we can see that: i) The proposed ResSCNN achieves the SOTA performance among the existing NR metrics and even outperforms some of the FR metrics. ii) The FR metrics outperform the NR metrics on the whole, which is in fact expected, because the lack of reference information in NR metrics increases the difficulty for PCQA task. But the performance of the proposed ResSCNN is not distant from that of the existing FR metrics. For example, if we compare ResSCNN and the FR metrics with best performance in three datasets, we have (0.60, 0.62) for ResSCNN vs. (0.70, 0.67) for MPED on LS-PCQA, (0.86, 0.81) for ResSCNN vs. (0.91, 0.89) for GraphSIM on SJTU-PCQA, and (0.72, 0.75) for ResSCNN vs. (0.74, 0.75) for PCQM and GraphSIM on WPC2.0. iii) The large-scale dataset brings more challenges to PCQA task. We can notice that both FR and NR metrics for PCQA have potentials for further improvement.

\par In summary, the proposed ResSCNN offers robust and competitive performance over three datasets, compared with the existing NR and even some FR metrics.

\subsection{Efﬁciency of The Proposed Dataset}

\par The built large-scale dataset can benefit the training of learning-based NR-PCQA metrics. We conduct an experiment to verify the improvements due to pre-training on the established LS-PCQA. The proposed ResSCNN is first pre-trained on the training set of LS-PCQA, then trained using the training set of SJTU-PCQA and tested on the testing set of WPC2.0. Next, the training database and testing database are switched and the experiment is repeated. The experiment results are listed in Table~\ref{cross1} and Table~\ref{cross2}.

\begin{table}[h]
  \centering
  \caption{Improvement of the built large-scale dataset for learning-based NR-PCQA metrics (Training on SJTU-PCQA and testing on WPC2.0 with pre-training on LS-PCQA and without).}
  \begin{footnotesize}
    \begin{tabular}{l|rr}
    \hline
          & \multicolumn{1}{l}{PLCC} & \multicolumn{1}{l}{SROCC} \\
    \hline
    ResSCNN & 0.47   & 0.48 \\
    \hline
    ResSCNN + LS-PCQA & 0.66   & 0.64 \\
    \hline
    \end{tabular}%
    \end{footnotesize}
  \label{cross1}%
\end{table}%

\begin{table}[h]
  \centering
  \caption{Improvement of the built large-scale dataset for learning-based NR-PCQA metrics (Training on WPC2.0 and testing on SJTU-PCQA with pre-training on LS-PCQA and without).}
  \begin{footnotesize}
    \begin{tabular}{l|rr}
    \hline
          & \multicolumn{1}{l}{PLCC} & \multicolumn{1}{l}{SROCC} \\
    \hline
    ResSCNN & 0.67   & 0.60 \\
    \hline
    ResSCNN + LS-PCQA & 0.75   & 0.69 \\
    \hline
    \end{tabular}%
    \end{footnotesize}
  \label{cross2}%
\end{table}%

\par We can see from Table~\ref{cross1} and Table~\ref{cross2} that the generalization capability has been improved due to pre-training on LS-PCQA. After being pre-trained on LS-PCQA, ResSCNN that is trained over SJTU-PCQA and tested on WPC2.0 achieves up to about 40\% and 33\% increase in PLCC and SROCC. ResSCNN that is trained over WPC2.0 and tested on SJTU-PCQA achieves up to about 12\% and 15\% increase in PLCC and SROCC as well.

\par In summary, the established LS-PCQA can benefit the learning-based NR-PCQA metrics, and enhances the generalization ability for the NR-PCQA task.

\subsection{Effect of Sampling}

\par Many 3D applications adopt dimensionality reduction techniques as a pre-processing method for input normalization, such as key point extraction and down-sampling. However, these dimensionality reduction techniques will introduce extra geometrical distortions into the point clouds, and thus should not be used in NR-PCQA task.
\par In this subsection, we conduct experiments to demonstrate that the dimensionality reduction techniques decrease the performance of NR-PCQA metrics. Specifically, we down-sample the point clouds to 400,000, 100,000, 50,000, 10,000 and 2,500 points respectively. The overall performance and performance for down-sampling distortion for ResSCNN are shown in Table~\ref{samplingeffect}.

\begin{table}[htbp]
  \centering
  \caption{Performance of ResSCNN with different sampling versions.}
  \resizebox{\linewidth}{!}{
   \begin{tabular}{l|rr|rr|rr|rr|rr|rr}
    \hline
          & \multicolumn{2}{c|}{\multirow{2}[0]{*}{Keeping all points}} & \multicolumn{10}{c}{Sampling to } \\
\cline{4-13}          & \multicolumn{2}{c|}{} & \multicolumn{2}{c|}{400,000 points} & \multicolumn{2}{c|}{100,000 points} & \multicolumn{2}{c|}{50,000 points} & \multicolumn{2}{c|}{10,000 points} & \multicolumn{2}{c}{2,500 points} \\
\cline{2-13}          & \multicolumn{1}{l}{PLCC} & \multicolumn{1}{l|}{SROCC} & \multicolumn{1}{l}{PLCC} & \multicolumn{1}{l|}{SROCC} & \multicolumn{1}{l}{PLCC} & \multicolumn{1}{l|}{SROCC} & \multicolumn{1}{l}{PLCC} & \multicolumn{1}{l|}{SROCC} & \multicolumn{1}{l}{PLCC} & \multicolumn{1}{l|}{SROCC} & \multicolumn{1}{l}{PLCC} & \multicolumn{1}{l}{SROCC} \\
\hline
    Overall & 0.60  & 0.62  & 0.53  & 0.56  & 0.50   & 0.54  & 0.50  & 0.51  & 0.42  & 0.44  & 0.44  & 0.43  \\
    Down-sampling distortion & 0.81  & 0.73  & 0.55   & 0.56  & 0.53  & 0.48  & 0.34  & 0.33  & 0.38  & 0.30  & 0.25  & 0.19  \\
    \hline
    \end{tabular}%
    }
  \label{samplingeffect}%
\end{table}%

\par We can see from Table~\ref{samplingeffect} that: i) Generally, the proposed ResSCNN tends to provide better performance with more points. ii) The dimensionality reduction techniques which remove part of the points will affect the results of NR-PCQA, which also corroborates our observations that NR-PCQA is different from other vision tasks and taking the complete samples into consideration is necessary.

\par Thus, the proposed ResSCNN should take entire point clouds as input to avoid introducing additional distortions in the pre-processing stage.

\subsection{Effectiveness of Hierarchical Features}

\par Hierarchical features have been applied in many image applications. The reason is that the semantic information of CNN from the shallow to deep layers has specific characteristics~\cite{Bai2020MFT}. The shallow features provide rich details of color and texture while deep features contain more conceptual and semantic information. Thus, it could be expected that the hierarchical features in PCQA task can improve the robustness under various distortions.
\par In this subsection, we conduct experiments on LS-PCQA to verify the rationality of the proposed hierarchical features. Specifically, we use the features from the first block in Fig. \ref{framework} as shallow features, and the features from the last block as deep features. The results are shown in Table~\ref{ShallowAndDeep}.

\begin{table}[htbp]
  \centering
  \caption{Performance using features from different depths. The proposed hierarchical feature provides the best overall performance.}
  \begin{footnotesize}
    \begin{tabular}{l|rr|rr|rr}
    \hline
    \multirow{2}[4]{*}{} & \multicolumn{2}{c|}{Shallow Feature} & \multicolumn{2}{c|}{Deep Feature} & \multicolumn{2}{c}{Hierarchical Feature} \\
\cline{2-7}          & \multicolumn{1}{l}{PLCC} & \multicolumn{1}{l|}{SROCC} & \multicolumn{1}{l}{PLCC} & \multicolumn{1}{l|}{SROCC} & \multicolumn{1}{l}{PLCC} & \multicolumn{1}{l}{\hfill SROCC} \\
    \hline
    ResSCNN on LS-PCQA &0.51&0.53&0.41&0.42&0.60&0.62  \\
    \hline
    \end{tabular}%
    \end{footnotesize}
  \label{ShallowAndDeep}%
\end{table}%

\par We can see from Table~\ref{ShallowAndDeep} that: i) The hierarchical feature outperforms the shallow and deep features alone. ii) Most of the distortion types in the built dataset are detailed distortions, and the shallow features provide rich details of color and texture. Thus, the shallow features can better handle detailed distortions. However, the conceptual and semantic distortions are also included in the PCQA dataset, such as the contrast distortion and luminance distortion, and the deep features should also need to be considered in NR-PCQA metrics.

\par In other words, the proposed hierarchical feature exhibits more robust performance.

\subsection{Effect of Network Depth}

\par In this part, we conduct experiments to investigate the impact of network depth on quality prediction over the LS-PCQA dataset. The results are shown in Table~\ref{networkdepth}.

\begin{table}[htbp]
  \centering
  \caption{Performance of ResSCNN with different depths. 1 block contains three sparse convolutional layers. The 4-blocks depth presents the best overall performance.}
  \begin{footnotesize}
    \begin{tabular}{l|r|r|r|r|r}
    \hline
          & \multicolumn{1}{l|}{1 block} & \multicolumn{1}{l|}{2 blocks} & \multicolumn{1}{l|}{3 blocks} & \multicolumn{1}{l|}{4 blocks} & \multicolumn{1}{l}{5 blocks} \\
    \hline
    PLCC  & 0.51  &  0.57     &   0.57    & \textbf{0.60}  & 0.46 \\
    SROCC & 0.53  &  0.59     &   0.58    & \textbf{0.62}  & 0.47 \\
    \hline
    \end{tabular}%
    \end{footnotesize}
  \label{networkdepth}%
\end{table}%

\par We can see from Table~\ref{networkdepth} that: i) The shallow feature has a good response to a large number of detailed and visually distorted samples in the dataset. But using only the shallow features cannot describe the dataset adequately. ii) The model tackles the dataset better with increasing depth of the network. However, the increase of the number of parameters may result in over-fitting. That is why we observe that the network with 4 blocks yields the best performance.

\subsection{Effectiveness of the Residual Module}

\par The proposed ResSCNN aims to handle point clouds with arbitrary number of points, and the pooling operation is adopted for the normalization of obtained features. Some part of the information in the features is partially damaged in this process. To enhance the expressiveness of output features, we compensate for the information loss using the residual connections.
\par
\begin{figure}[htbp]
	\centering
	\includegraphics[width=0.7\linewidth]{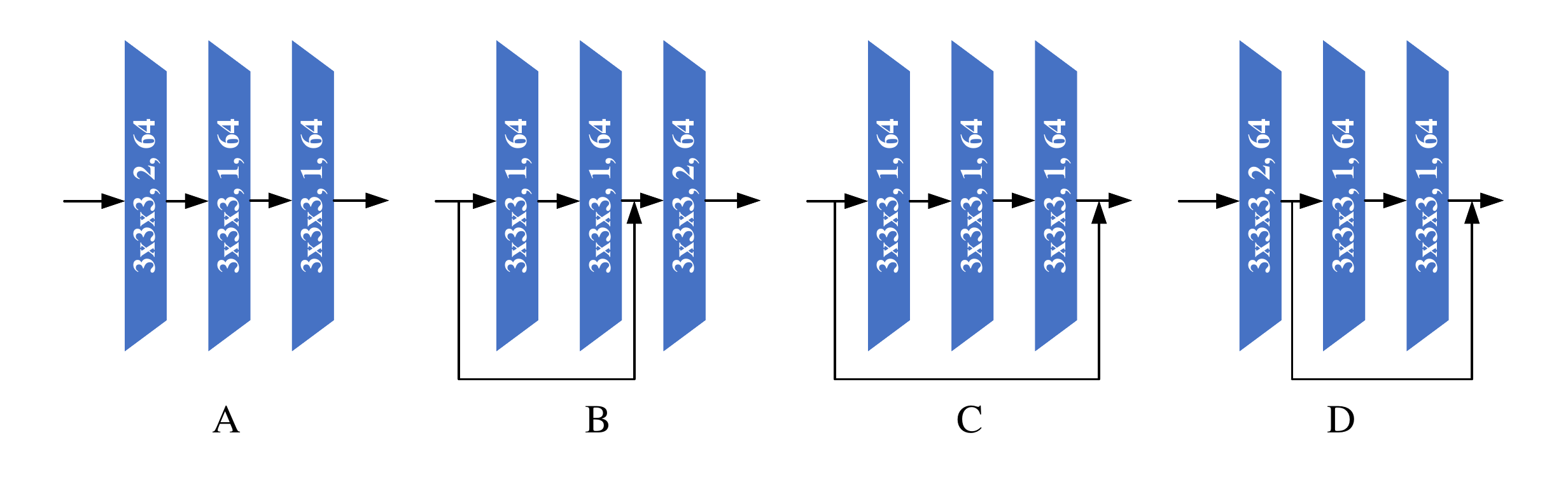}
	\caption{Alternatives of the residual connections.}
	\label{otherresidual}

\end{figure}

 In this part, we conduct experiments to illustrate the effectiveness of the residual module. Specifically, we compare the prediction accuracy of ResSCNN on LS-PCQA with several alternatives of residual connections. The testing networks are composed of four identical blocks as shown in Fig. \ref{otherresidual}. For dimensionality matching, the first blocks in residual networks B-D have residual connection spanning 2 and 3 layers. The results are shown in Table~\ref{reseffect}.

\begin{table}[htbp]
  \centering
  \caption{Performance of ResSCNN with different residual connection schemes.}
  \begin{footnotesize}
    \begin{tabular}{l|rr}
    \hline
          & \multicolumn{1}{l}{PLCC} & \multicolumn{1}{l}{SROCC} \\
    \hline
    A: Without residual connection & 0.56  & 0.56  \\
    B: 1, 2-layers residual connection  & 0.57  & 0.57  \\
    C: 1, 3-layers residual connection  & 0.59  & 0.60  \\
    D: 2, 3-layers residual connection  & 0.60  & 0.62  \\
    \hline
    \end{tabular}%
    \end{footnotesize}
  \label{reseffect}%
\end{table}%

\par We can see from Table~\ref{reseffect} that the use of residual connections improves the accuracy of quality prediction. As a result, the final design of our proposed network shown in Fig. \ref{framework} does adopt residual connections.

\subsection{Performance for Different Distortions}

\par In this subsection, we evaluate the performance of several FR-PCQA metrics, including GraphSIM and PCQM, together with NR-PCQA metrics, including PQA-Net, IT-PCQA and ResSCNN, under different distortions. Specifically, on the basis of the experiment results in Section~\ref{sec:LSexperiment}, the PLCC and SROCC of these PCQA metrics are calculated for each distortion type. The results are listed in Table~\ref{alldistortions}, and the best results for FR and NR metrics are highlighted in bold respectively.

\begin{table}[htbp]
  \centering
  \caption{Performance of FR and NR PCQA metrics for different distortion types on LS-PCQA.}
  \begin{tiny}
    \begin{tabular}{p{13.5em}|rr|rr|rr|rr|rr}
    \hline
    \multicolumn{1}{r|}{} & \multicolumn{2}{c|}{GraphSIM (FR)} & \multicolumn{2}{c|}{MPED (FR)} & \multicolumn{2}{c|}{PQA-Net (NR)} & \multicolumn{2}{c|}{IT-PCQA (NR)} & \multicolumn{2}{c}{ResSCNN (NR)} \\
    \hline
    \multicolumn{1}{l|}{Distortion} & \multicolumn{1}{l}{PLCC} & \multicolumn{1}{l|}{SROCC} & \multicolumn{1}{l}{PLCC} & \multicolumn{1}{l|}{SROCC} & \multicolumn{1}{l}{PLCC} & \multicolumn{1}{l|}{SROCC} & \multicolumn{1}{l}{PLCC} & \multicolumn{1}{l|}{SROCC} & \multicolumn{1}{l}{PLCC} & \multicolumn{1}{l}{SROCC} \\
    \hline
    \multicolumn{1}{l|}{ColorNoise} & \textbf{0.87}  & \textbf{0.83}  & 0.84  & 0.82  & 0.52  & 0.39  & 0.47  & 0.21  & \textbf{0.64}  & \textbf{0.66}  \\
    \multicolumn{1}{l|}{ColorQuantizationDither} & \textbf{0.63}  & \textbf{0.55}  & 0.60  & 0.47  & 0.62  & 0.27  & 0.64  & 0.21  & \textbf{0.67}  & \textbf{0.53}  \\
    \multicolumn{1}{l|}{ContrastDistortion} & \textbf{0.72}  & \textbf{0.64}  & 0.67  & 0.51  & 0.57  & 0.22  & 0.22  & -0.16  & \textbf{0.62}  & \textbf{0.36}  \\
    \multicolumn{1}{l|}{CorrelatedGaussianNoise} & \textbf{0.81}  & \textbf{0.76}  & 0.66  & 0.60  & 0.81  & 0.74  & 0.30  & 0.22  & \textbf{0.86}  & \textbf{0.83}  \\
    \multicolumn{1}{l|}{DownSample} & \textbf{0.97}  & \textbf{0.97}  & \textbf{0.97}  & 0.96  & 0.52  & 0.38  & 0.38  & 0.21  & \textbf{0.81}  & \textbf{0.73}  \\
    \multicolumn{1}{l|}{GammaNoise} & \textbf{0.59}  & \textbf{0.46}  & 0.53  & 0.38  & 0.34  & 0.09  & 0.34  & -0.18  & \textbf{0.40}  & \textbf{0.38}  \\
    \multicolumn{1}{l|}{GaussianNoise} & 0.55  & \textbf{0.49}  & \textbf{0.56}  & 0.31  & 0.69  & 0.65  & 0.75  & 0.53  & \textbf{0.81}  & \textbf{0.83}  \\
    \multicolumn{1}{l|}{GaussianShifting} & \textbf{0.95}  & \textbf{0.86}  & 0.89  & 0.78  & 0.21  & -0.03  & 0.51  & -0.48  & \textbf{0.81}  & \textbf{0.78}  \\
    \multicolumn{1}{l|}{HighFrequencyNoise} & \textbf{0.67}  & \textbf{0.76}  & 0.58  & 0.61  & 0.77  & 0.71  & 0.48  & 0.18  & \textbf{0.96}  & \textbf{0.92}  \\
    \multicolumn{1}{l|}{LocalMissing} & \textbf{1.00}  & \textbf{1.00}  & 0.70  & 0.85  & 0.84  & 0.68  & 0.19  & -0.26  & \textbf{0.91}  & \textbf{0.91}  \\
    \multicolumn{1}{l|}{LocalOffset} & \textbf{0.93}  & 0.91  & 0.47  & \textbf{0.93}  & 0.59  & 0.46  & 0.44  & -0.22  & \textbf{0.80}  & \textbf{0.79}  \\
    \multicolumn{1}{l|}{LocalRotation} & \textbf{0.88}  & \textbf{0.77}  & 0.43  & 0.71  & 0.72  & 0.68  & 0.41  & -0.19  & \textbf{0.79}  & \textbf{0.70}  \\
    \multicolumn{1}{l|}{LumaNoise} & \textbf{0.77}  & \textbf{0.74}  & \textbf{0.77}  & 0.67  & 0.59  & 0.37  & 0.40  & -0.30  & \textbf{0.78}  & \textbf{0.51}  \\
    \multicolumn{1}{l|}{MeanShift} & \textbf{0.78}  & \textbf{0.74}  & 0.77  & 0.67  & 0.40  & 0.22  & 0.39  & -0.30  & \textbf{0.79}  & \textbf{0.46}  \\
    \multicolumn{1}{l|}{MultiplicativeGaussianNoise} & \textbf{0.72}  & \textbf{0.65}  & 0.63  & 0.60  & 0.80  & 0.82  & 0.53  & 0.44  & \textbf{0.85}  & \textbf{0.83}  \\
    \multicolumn{1}{l|}{PoissonNoise} & \textbf{0.74}  & 0.58  & 0.72  & \textbf{0.67}  & 0.45  & -0.06  & 0.04  & 0.08  & \textbf{0.65}  & \textbf{0.41}  \\
    \multicolumn{1}{l|}{QuantizationNoise} & \textbf{0.76}  & \textbf{0.47}  & 0.68  & 0.42  & \textbf{0.59}  & 0.32  & 0.46  & 0.26  & 0.53  & \textbf{0.39}  \\
    \multicolumn{1}{l|}{RayleighNoise} & \textbf{0.75}  & \textbf{0.60}  & 0.72  & \textbf{0.60}  & \textbf{0.74}  & 0.74  & 0.31  & 0.22  & 0.73  & \textbf{0.76}  \\
    \multicolumn{1}{l|}{SaltpepperNoise} & \textbf{0.90}  & \textbf{0.83}  & 0.89  & 0.81  & 0.60  & 0.57  & 0.47  & 0.30  & \textbf{0.66}  & \textbf{0.65}  \\
    \multicolumn{1}{l|}{SaturationDistortion} & \textbf{0.50}  & 0.42  & 0.44  & \textbf{0.44}  & \textbf{0.77}  & \textbf{0.75}  & 0.52  & 0.16  & 0.53  & 0.33  \\
    \multicolumn{1}{l|}{UniformNoise} & \textbf{0.79}  & \textbf{0.77}  & 0.64  & 0.66  & \textbf{0.80}  & 0.59  & 0.46  & 0.19  & 0.67  & \textbf{0.65}  \\
    \multicolumn{1}{l|}{UniformShifting} & 0.84  & 0.80  & \textbf{0.93}  & \textbf{0.92}  & 0.40  & -0.14  & 0.65  & -0.73  & \textbf{0.85}  & \textbf{0.74}  \\
    \multicolumn{1}{l|}{VPCC\_lossy-geom-lossy-attrs} & \textbf{1.00}  & \textbf{1.00}  & 0.77  & 0.68  & 0.67  & 0.22  & 0.56  & 0.23  & \textbf{0.80}  & \textbf{0.71}  \\
    \multicolumn{1}{l|}{AVS\_limitlossyG-lossyA} & \textbf{0.99}  & 0.84  & 0.98  & \textbf{0.87}  & \textbf{0.95}  & 0.73  & 0.90  & -0.56  & 0.91  & \textbf{0.75}  \\
    \multicolumn{1}{l|}{AVS\_losslessG-limitlossyA} & \textbf{0.88}  & \textbf{0.88}  & 0.83  & 0.85  & 0.35  & -0.04  & 0.35  & -0.16  & \textbf{0.37}  & \textbf{0.34}  \\
    \multicolumn{1}{l|}{AVS\_losslessG-lossyA} & \textbf{0.87}  & \textbf{0.85}  & 0.86  & \textbf{0.85}  & 0.51  & 0.30  & \textbf{0.74}  & \textbf{0.49}  & 0.45  & 0.46  \\
    \multicolumn{1}{l|}{GPCC\_lossless-geom-lossy-attrs} & \textbf{0.82}  & \textbf{0.89}  & 0.54  & 0.60  & 0.37  & -0.01  & \textbf{0.62}  & \textbf{0.25}  & 0.41  & 0.20  \\
    \multicolumn{1}{l|}{GPCC\_lossless-geom-nearlossless-attrs} & \textbf{0.94}  & \textbf{0.99}  & 0.90  & 0.85  & 0.34  & -0.13  & \textbf{0.85}  & \textbf{0.62}  & 0.72  & 0.47  \\
    \multicolumn{1}{l|}{GPCC\_lossy-geom-lossy-attrs} & \textbf{0.98}  & \textbf{0.94}  & 0.95  & 0.89  & 0.84  & 0.56  & \textbf{0.85}  & -0.64  & \textbf{0.85}  & \textbf{0.65}  \\
    \multicolumn{1}{l|}{Octree} & \textbf{0.89}  & \textbf{0.83}  & 0.80  & 0.75  & \textbf{0.78}  & \textbf{0.65}  & 0.64  & -0.39  & 0.38  & 0.27  \\
    PossionReconstruction & 0.69  & 0.32  & \textbf{0.97}  & \textbf{0.95}  & \textbf{0.72}  & 0.08  & 0.52  & \textbf{0.32}  & 0.62  & 0.21  \\
    \hline
    \end{tabular}%
  \end{tiny}
  \label{alldistortions}%
\end{table}%

\par We can see from Table~\ref{alldistortions} that: i) Compared with the existing projection-based NR-PCQA metrics, the proposed ResSCNN exhibits the best performance for almost all distortion types. Specifically, the proposed ResSCNN performs better in some geometrical distortions, such as the Local Missing distortion. The reason is that the projection renders some geometrical loss difficult to detect. ii) All the NR-PCQA metrics exhibit poor performance in some color distortions, such as the Contrast Distortion and Poisson Noise. Compared with the geometrical distortion, these color distortions are hard to be perceived without the assistance of reference. We hypothesize that extra semantic features in terms of color distributions need to be accounted for in the design of NR-PCQA. iii) The FR-PCQA metrics are more robust than the NR metrics. With the original point clouds as references, the FR-PCQA metrics are easier to quantify the influence of distortions on human perception. Besides, the sensitivity of FR metrics is different. For example, GraphSIM is more sensitive to GPCC distortions, while MPED is more sensitive to the reconstruction distortion.

\par On the whole, the proposed ResSCNN exhibits the best performance for almost all distortion types compared with the existing NR-PCQA metrics. Moreover, the PCQA becomes more challenging without references, and the learning-based NR-PCQA metrics have potentials for further improvement.

\section{Conclusion}\label{sec:conclusion}

\par In this work, we proposed a NR-PCQA metric, ResSCNN. To meet the requirement of data scale for training learning-based metrics, we firstly built a PCQA dataset of the largest size at present. The built LS-PCQA dataset contains more than 22,000 distorted samples derived from 104 original reference point clouds with 31 impairment types at 7 distortion levels. Leveraging the newly built dataset, a NR-PCQA metric based on sparse CNN was proposed. Experiment results have demonstrated the efficiency of our proposed ResSCNN which achieves the SOTA performance among the existing NR-PCQA metrics and is even competitive compared with the FR-PCQA metrics. Besides, we have proved that the proposed large-scale dataset can help improve the generalization ability of the learning-based NR-PCQA metrics.

\bibliographystyle{ACM-Reference-Format}
\bibliography{ref}

\appendix

\section{Details for Distortion Types in LS-PCQA Dataset}\label{sec:distortiondetails}

\par The details for each distortion type in the built dataset are given in the following.

\begin{enumerate}
\item {\sf Color noise~\cite{Ponomarenko2015TID2013,Yang2020TMM3DTO2D}}: Color noise is applied to the photometric attributes of the points. We inject the noise for  $10\%$,$20\%$, $30\%$, $40\%$, $50\%$, $60\%$, and $70\%$ points that are randomly selected, where noise levels are respectively and again randomly given within  $\pm10$,$\pm20$, $\pm30$, $\pm40$, $\pm50$, $\pm60$, and $\pm70$ for corresponding points (e.g., $10\%$ random points with $\pm10$ noise,  $30\%$ random points with $\pm30$, and so on so forth). Noise is equally applied to R, G, B attributes. Cropping is used here and other distortions if the noisy intensity, $\tilde{p} = p + n $, is out of the range of [0, 255], i.e., if $\tilde{p} < 0$, $\tilde{p} = 0$; and if $\tilde{p} > 255$, $\tilde{p} = 255$.
  \item {\sf Additive Gaussian noise~\cite{Boyat2015ImageNoise,Ponomarenko2015TID2013,Yang2020TMM3DTO2D}}: The photometric attributes are added with Gaussian noise with different intensities. The signal to noise ratios for different distortion levels are respectively 13dB, 11dB, 9dB, 7dB, 5dB, 3dB, and 1dB.
  \item {\sf High-frequency noise~\cite{Ponomarenko2015TID2013}}: Gaussian noise is added to the high-frequency part of the photometric attributes of point clouds. The variances of the added zero-mean Gaussian noise are 0.001, 0.003, 0.005, 0.0075, 0.01, 0.03, and 0.05 respectively by the function imnoise() in Matlab.
  \item {\sf Quantization noise~\cite{Boyat2015ImageNoise,Ponomarenko2015TID2013}}: This distortion derives from the color quantization. To implement the quantization noise, we equally quantify the photometric attributes. For different distortion levels, we set the quantization step to 27, 33, 39, 47, 55, 65, and 76 respectively.
  \item {\sf Mean shift (intensity shift)~\cite{Ponomarenko2015TID2013}}: The mean shift is applied to the photometric attributes of the point clouds. The photometric attributes are respectively enhanced 10, 20, 30, 40, 50, 60, and 70.
  \item {\sf Contrast change~\cite{Ponomarenko2015TID2013}}: The contrast change is to contaminate the photometric attributes of point clouds in perceptual contrast. To implement the contrast change, the R, G, B attributes are 1.1, 1.2, 1.3, 1.4, 1.5, 1.6, and 1.7 power of original values respectively.
  \item {\sf Change of color saturation~\cite{Ponomarenko2015TID2013}}:  The color saturation distortion is to contaminate the photometric attributes of point clouds in color saturation. The saturation of photometric attributes is computed based on lightness. Then the increment of saturation is set to $-10\%$, $-25\%$, $-40\%$, $-55\%$, $-70\%$, $-85\%$, and $-100\%$ respectively.
  \item {\sf Spatially correlated noise~\cite{Ponomarenko2015TID2013}}: The photometric attributes of point clouds are added to the zero-mean correlated Gaussian noise whose standard deviation is 10, 20, 30, 40, 50, 60, and 70 respectively.
  \item {\sf Multiplicative Gaussian noise~\cite{Ponomarenko2015TID2013}}: To implement the multiplicative Gaussian noise, the photometric attributes of point clouds are multiplied by the Gaussian white noise with different intensities. The intensities of multiplied Gaussian noise are set to 1, 3, 5.5, 8, 10.5, 13, and 15.5 ($\times$1e-4) dBW.
  \item {\sf Color quantization with dither~\cite{Boyat2015ImageNoise,Ponomarenko2015TID2013}}: This distortion is produced by the rgb2ind operation. To implement the color quantization with dither distortion, we convert the true R, G, B color to index color. The index number of colors for different distortion levels is set to 24, 16, 12, 8, 6, 4, and 2.
  \item {\sf Down-sampling~\cite{Ponomarenko2015TID2013,Yang2020TMM3DTO2D}}: The down-sampling distortion is caused by the missing points. We randomly down-sample the point clouds by removing $15\%$, $30\%$, $45\%$, $60\%$, $70\%$, $80\%$, $90\%$ points from the original point clouds.
  \item {\sf Salt-pepper noise~\cite{Boyat2015ImageNoise,Ponomarenko2015TID2013}}: The salt-and-pepper noise is the white or black spots that appear randomly in the photometric attributes of point clouds. To produce the salt-and-pepper noise for different distortion levels, we set the proportion of random points where the salt-and-pepper noise emerges to 0.02, 0.05, 0.1, 0.15, 0.2, 0.25, and 0.30 respectively.
  \item {\sf Rayleigh noise~\cite{Boyat2015ImageNoise}}: The Rayleigh noise is equally added to all the photometric attributes. We use the function raylrnd() in Matlab to generate the Rayleigh noise. For different distortion levels, parameter $B$ of raylrnd() is set to 10, 20, 30, 40, 50, 60, and 70 respectively.
  \item {\sf Gamma noise~\cite{Boyat2015ImageNoise}}: Gamma noise obeys the Gamma curve distribution. To realize the Gamma noise, $B$ noises which obey exponential distribution are superimposed, formulated as \[E = \sum\limits_{i = 1}^B {{E_i}}  = \sum\limits_{i = 1}^B { - \frac{1}{a}In[1 - U(0,1)]} \]
      In this work, $B$ is set to 3 and for different distortion levels, $a$ is set to 0.1, 0.08, 0.07, 0.06, 0.05, 0.04, and 0.03 respectively.
  \item {\sf Uniform noise (white noise)~\cite{Boyat2015ImageNoise,Ponomarenko2015TID2013}}: The white noise is added to the photometric attributes of point clouds. To implement this distortion, we randomly offset R, G, B values of each point among [-10,10], [-20,20], [-30,30], [-40,40], [-50,50], [-60,60], and [-70,70] respectively.
  \item {\sf Poisson noise~\cite{Boyat2015ImageNoise}}: The Poisson noise is equally added to all the photometric attributes. We use the function poissrnd() in Matlab to generate the Poisson noise. For different distortion levels, parameter $B$ of poissrnd() is set to 10, 20, 30, 40, 50, 60, and 70 respectively.
  \item {\sf Gaussian geometry shifting}: We apply Gaussian distributed geometric shift to each point randomly. In this work, all the points will be superimposed with a zero-mean Gaussian geometry shift whose standard deviation is $0.1\%$, $0.25\%$, $0.4\%$, $0.55\%$, $0.7\%$, $0.85\%$, $1\%$ of the bounding box respectively.
  \item {\sf Uniform geometry shifting}: Uniform geometry shifting is applied to the geometry attributes of point clouds. The shifting are applied in $10\%$,$20\%$, $30\%$, $40\%$, $50\%$, $60\%$, and $70\%$ points that are randomly selected, where the shifting ranges are respectively $\pm0.5\%$,$\pm1\%$, $\pm2\%$, $\pm3\%$, $\pm4\%$, $\pm5\%$, and $\pm7\%$ of the bounding box for corresponding points (e.g., $10\%$ random points with $\pm0.5\%$ noise,  $30\%$ random points with $\pm2\%$, and so on so forth).
  \item {\sf Local Missing}: This distortion derives from the missing of local patches. To implement this distortion, we define the space anchor with the size of 0.3 times the bounding box. With the increase of distortion level, the additional 1, 1, 2, 2, 3, 3, 4 anchors are selected for each distortion level on the basis of the selected anchors in the previous distortion level (e.g., after $1+1$ space anchors are selected for distortion of level 2, another 2 different anchors are selected for distortion of level 3). Then the points in the selected anchors are removed.
  \item {\sf Local offset}: This distortion derives from the local dislocation of point clouds. To implement this distortion, the selection of anchors is as same as that in the local missing distortion, and for each distortion level, the values of geometry attributes for each point in the selected anchors will increase by 5\% of the maximum side length of the bounding box.
  \item {\sf Local rotation}: This distortion derives from the local warp of point clouds. To implement this distortion, the selection of anchors is as same as that in the local missing distortion, and for each distortion level, the points in the selected anchors will be rotated 20, 25, 30, 35, 40, 45, and 50 degrees respectively along the x-axis.
  \item {\sf Luminance distortion~\cite{Ponomarenko2015TID2013}}: Luminance distortion is to contaminate the point clouds in the luminance component. To implement the luminance distortion, the photometric attributes of point clouds are converted from RGB space to YCbCr space. Then the luminance component is added to an offset which is 20, 50, 70, 90, 110, 130, and 150 respectively.
  \item {\sf Poisson Reconstruction~\cite{Michael2006PoissonRC}}: The reconstruction distortion comes from the impairment between the original and constructed point clouds. To implement this distortion, we conduct the Poisson surface reconstruction to convert the down-sampled point cloud to mesh. Then the face of mesh is randomly sampled to restore the number of points before down-sampling. The proportion of down-sampling is as same as that in the down-sampling distortion.
    \item {\sf Octree Compression~\cite{Yang2020TMM3DTO2D}}: This distortion comes from the octree compression provided in well-known Point Cloud Library (PCL). Here, we have experimented different compression levels by setting octree resolution at 8, 10, 12, 14, 16, 18 and 20 respectively.
  \item {\sf MPEG GPCC with lossless geometry and lossy attributes}: This distortion derives from the GPCC compression under lossless geometry and lossy attributes (test condition C1) coding in all-intra mode using mpeg-pcc-tmc13 v11.0 (octree-predlift). To implement this distortion, we set the parameter $qp$ to 27, 31, 35, 39, 43, 47, and 51 respectively.
  \item {\sf MPEG GPCC with lossless geometry and nearlossless attributes}: This distortion derives from the GPCC compression under lossless geometry and nearlossless attributes (test condition CY) coding in all-intra mode using mpeg-pcc-tmc13 v11.0 (octree-predlift). To implement this distortion, we set the parameter $qp$ to 10, 16, 22, 28, 34, 40, and 46 respectively.
  \item {\sf MPEG GPCC with lossy geometry and lossy attributes}: This distortion derives from the GPCC compression under lossy geometry and lossy attributes (test condition C2) coding in all-intra mode using mpeg-pcc-tmc13 v11.0 (octree-predlift). To implement this distortion, we set the parameter $positionQuantizationScale$ to 0.9375, 0.875, 0.75, 0.5, 0.25, 0.125, 0.0625 and set $qp$ to 27, 31, 35, 39, 43, 47, 51 respectively.

  \item {\sf MPEG VPCC with lossy geometry and lossy attributes}:  This distortion derives from the VPCC compression under lossy geometry and lossy attributes (test condition C2) coding in all-intra mode using mpeg-pcc-tmc2 v11.0. To implement this distortion, we set the parameter $geometryQP$ to 16, 20, 24, 28, 32, 36, 40 and set $textureQP$ to 22, 27, 32, 37, 42, 47, 51 respectively.
  \item {\sf AVS PCRM with limitlossy geometry and lossy attributes}: This distortion derives from the AVS compression under limitlossy-geometry-lossy-attributes coding in all-intra mode using avs-pcc-pcrm v1.0. To implement this distortion, we set the parameter $geom\_quant\_step$ to 1.14286, 1.33333, 2, 4, 8, 12, 16 and set $attr\_quant\_param$ to 8, 16, 24, 32, 40, 44, 48 respectively.
  \item {\sf AVS PCRM with lossless geometry and limitlossy attribute}: This distortion derives from the AVS compression under lossless-geometry-limitlossy-attributes coding in all-intra mode using avs-pcc-pcrm v1.0. To implement this distortion, we set the parameter $attr\_quant\_param$ to 8, 16, 24, 32, 40, 44, 48 respectively.
  \item {\sf AVS PCRM with lossless geometry and lossy attributes}: This distortion derives from the AVS compression under lossless-geometry-lossy-attributes coding in all-intra mode using avs-pcc-pcrm v1.0. To implement this distortion, we set the parameter $attr\_quant\_param$ to 8, 16, 24, 32, 40, 44, 48 respectively.

\end{enumerate}

\end{document}